\documentclass[aps,prb,singlecolumn,superscriptaddress]{revtex4-2}

\usepackage{graphicx}
\usepackage{amsmath}
\usepackage{amssymb}
\usepackage{mathtools}
\usepackage{xcolor}
\usepackage[colorlinks=true, citecolor=blue, urlcolor=blue, linkcolor=blue,bookmarks=false,hypertexnames=true]{hyperref}

\usepackage{array}
\usepackage[column=O]{cellspace}
\newcolumntype{P}[1]{>{\centering\arraybackslash}p{#1}}

\makeatletter
\def\maketitle{
\@author@finish
\title@column\titleblock@produce
\suppressfloats[t]}
\makeatother

\begin{document}

\graphicspath{}
\setlength{\parskip}{\medskipamount}

\title{Valley-spin polarization at zero magnetic field induced by strong hole-hole interactions in monolayer WSe$_2$}


\author{Justin Boddison-Chouinard}
    \affiliation{Quantum and Nanotechnologies Research Centre, National Research Council of Canada, Ottawa, Ontario, K1A 0R6}
    \affiliation{Department of Physics, University of Ottawa, Ottawa, Ontario, K1N 9A7}
\author{Marek Korkusinski}
    \affiliation{Quantum and Nanotechnologies Research Centre, National Research Council of Canada, Ottawa, Ontario, K1A 0R6}
\author{Alex Bogan}
    \affiliation{Quantum and Nanotechnologies Research Centre, National Research Council of Canada, Ottawa, Ontario, K1A 0R6}
\author{Pedro Barrios}
    \affiliation{Quantum and Nanotechnologies Research Centre, National Research Council of Canada, Ottawa, Ontario, K1A 0R6}
\author{Philip Waldron}
    \affiliation{Quantum and Nanotechnologies Research Centre, National Research Council of Canada, Ottawa, Ontario, K1A 0R6}
\author{Kenji Watanabe}
    \affiliation{Research Center for Electronic and Optical Materials, National Institute for Materials Science, 1-1 Namiki, Tsukuba 305-0044, Japan}
\author{Takashi Taniguchi}
    \affiliation{Research Center for Materials Nanoarchitectonics, National Institute for Materials Science,  1-1 Namiki, Tsukuba 305-0044, Japan}
\author{Jaros\l{}aw Paw\l{}owski}
    \affiliation{Institute of Theoretical Physics, Wroc\l{}aw University of Science and Technology, Wroc\l{}aw, Poland}
\author{Daniel Miravet}
    \affiliation{Department of Physics, University of Ottawa, Ottawa, Ontario, K1N 9A7}
\author{Pawel Hawrylak}
    \affiliation{Department of Physics, University of Ottawa, Ottawa, Ontario, K1N 9A7}
\author{Adina Luican-Mayer}
    \affiliation{Department of Physics, University of Ottawa, Ottawa, Ontario, K1N 9A7}
    \affiliation{Materials Science Division, Argonne National Laboratory, Lemont, Illinois 60439, USA}
\author{Louis Gaudreau}
    \affiliation{Quantum and Nanotechnologies Research Centre, National Research Council of Canada, Ottawa, Ontario, K1A 0R6}

\begin{abstract}
Monolayer transition metal dichalcogenides have emerged as prominent candidates to explore the complex interplay between the spin and the valleys degrees of freedom. The strong spin-orbit interaction and broken inversion symmetry within these materials lead to the spin-valley locking effect, in which carriers occupying the K and K' valleys of the reciprocal space must have opposite spin depending on which valley they reside. This effect is particularly strong for holes due to a larger spin-orbit gap in the valence band. By reducing the dimensionality of a monolayer of tungsten diselenide to 1D via electrostatic confinement, we demonstrate that spin-valley locking in combination with strong hole-hole interactions lead to a ferromagnetic state in which hole transport through the 1D system is spin-valley polarized, even without an applied magnetic field, and that the persistence of this spin-valley polarized configuration can be tuned by a global back-gate.
 This observation opens the possibility of implementing a robust and stable valley polarized system, essential for valleytronic applications.

\end{abstract}

\maketitle

Charge, spin, and valley degrees of freedom are fundamental in condensed matter physics not only because they enable storing and processing information in devices but also because their interplay can lead to exotic phases of matter. Despite advancements, there are still challenges in understanding and manipulating these interactions, particularly in the context of quantum technologies and quantum confined device architectures, where they can be affected by quantization effects, modified coupling strengths or many body interactions.
Among material platforms, two-dimensional (2D) transition metal dichalcogenides (TMDs) emerged as particularly attractive for spintronics and valleytronics \cite{Wu2013,Schaibley2016,  Lee2016,  Vitale2018, Lin2018, Mrudul2021,Altintas2021, Rana2023,Pawlowski2024_2}. One of the most remarkable properties of monolayer and odd multilayer TMDs is that due to strong spin-orbit interaction and broken spatial inversion symmetry, the $K$ and $K'$ valleys of the Brillouin zone are spin-split at zero magnetic field\cite{Xiao2012}. This splitting is particularly strong for holes in the valence band reaching $\approx$ 450 meV\cite{Falko2015, Peeters2018}, which leads to robust spin-valley locking where spin-up ($\uparrow$) holes reside in the $K$ valley and spin-down ($\downarrow$) holes in the $K'$ valley. While early observations of spin-valley locking were achieved in 2D devices using optical probes \cite{Zeng2012,Mak2012,Scrace2015} and 
transport\cite{Dean2023, Feldman2024}, only more recent progress in device fabrication permitted addressing this phenomenon in quantum confined architectures \cite{Weber2023, Pisoni2017, Marinov2017,Epping2018,Sakanashi2021, JBC2023}, however many of the observed features remain elusive.

In this article, we combine experimental evidence with theoretical calculations and demonstrate the role that spin-valley locking plays in the properties of 1D transport in monolayer tungsten diselenide (WSe$_2$). We show how hole-hole interactions lead to spin-valley polarized transport at zero magnetic field that can be tuned by controlling the carrier density in the device. We show that in TMDs, the combination of spin-valley locking and strong exchange interactions are two ingredients that are sufficient to fully understand spin-valley polarization at zero magnetic field.

\begin{figure}[t]
    \centering
    \includegraphics[width=0.5\textwidth]{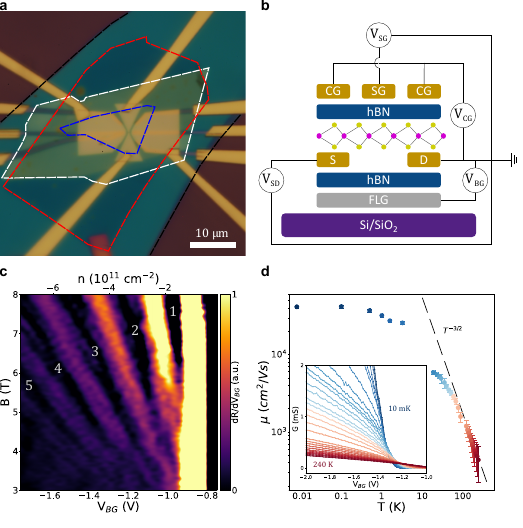}
    \caption{\textbf{Device architecture and electrical characterization.} \textbf {(a)} Optical micrograph of the WSe$_2$ device where the graphite (red), bottom hBN (black), monolayer WSe$_2$ (blue), and top hBN (white) are outlined. \textbf {(b)} Schematic illustration of the device architecture. \textbf {(c)} Landau fan diagram of the differential resistance $dR/dV_{BG}$ measured at 10 mK. The first five Landau levels are labelled with their respective filling factors. \textbf {(d)} Field-effect mobility as a function of temperature extracted from the activation curves plotted in the inset. A dashed line represents $\mu \propto T^{-3/2}$ as a guide, indicating that optical phonons are the dominating scattering mechanism at high temperatures\cite{Kaasbjerg2012}. The inset shows 4-probe conductance measurement as a function of the back-gate voltage when V$_{SG} = 0$ V and V$_{CG} = -7$ V.}
\end{figure}



In our experiments we use a device architecture having a monolayer WSe$_2$ encapsulated by hexagonal boron nitride (hBN), electrical contacts pre-patterned using chromium/platinum (Cr/Pt), a graphite back-gate, and accumulation and depletion top gates\cite{JBC2023}. Figure 1a shows an optical micrograph of the device and its cross-sectional view is represented in Fig.\,1b, where the top gates can be grouped into contact gates and split gates, labelled as CG and SG. The device contacts are activated by applying a negative voltage to the contact gates which locally increase the hole concentration in the contact region of the WSe$_2$. Applying a positive voltage on the split gates depletes the underlying region from holes and defines a tunable one-dimensional (1D) transport channel in the monolayer WSe$_2$ sheet with a lithographic width of 200 nm and length of 600 nm. The graphite back-gate tunes the carrier density in the entirety of the WSe$_2$ flake with the exception of the areas above the metal contacts due to screening. This results in total tunability of the carrier density without impacting the quality of the contacts. In this study, the contact gates were fixed to -7 V resulting in an average resistance per contact of approximately 2 k$\Omega$ at 10 mK (see SI for contact resistance details). Independent control of the carrier density of the contacts and of the active region enables the study of important transport properties in monolayer WSe$_2$, including quantum transport across a 1D channel.

To demonstrate the excellent electronic properties of  the device, we first perform Landau level spectroscopy experiments. Figure 1c shows the differential 4-point resistance dR/dV$_{BG}$ as a function of back-gate voltage (V$_{BG}$) and perpendicular magnetic field (B$_{\perp}$) at 10 mK. We observe Landau quantization at fields as low as 3 T, with the zeroth Landau level ($\nu = 1$, labelled as 1 in Fig.\,1c) emerging around 6 T. Low-filling factor integer quantum Hall states have been previously observed in monolayer TMDs using contactless methods such as electronic compressibility measurements\cite{Dean2018, Dean2020}, but have remained elusive in transport measurements until very recently\cite{Dean2023, Dean2023_2}, due to high contact resistances and low mobilities.

Further confirming the high quality of our device, Fig.\,1d shows the field-effect mobilities as a function of temperature, which have been calculated from activation curves (inset to Fig.\,1d) of our sample using the gate capacitance extracted from the Landau fan diagram in Fig.\,1c. At  240 K, we extract a field-effect mobility $\mu_{FE} =$ 400 $\pm$ 200 cm$^2$/(Vs). As the temperature decreases, the mobility increases following the expected power law, $\mu \propto T^{-\gamma}$, with $\gamma \approx 3/2$, indicating that at higher temperatures, optical phonons are the dominant scattering mechanism\cite{Kaasbjerg2012, Movva2015}. For lower temperatures, the mobility saturates and it is dominated by scattering impurities. At 10 mK, we achieve a mobility of 41 000 $\pm$ 2 000 cm$^2$/(Vs), comparable to recent reports of high-quality flux-grown monolayer WSe$_2$ samples\cite{Dean2023_2, Kim2024}.

We control the width of the 1D channel electrostatically by applying 4 V to the lower split gate and by tuning the voltage applied to the upper split gate (V$_{SG}$), resulting in the data presented in Fig.\,2a at B$_\perp$ = 0 T. The 4-terminal conductance $G$ is quantized, generating well-defined plateaus at multiples of $2G_0$ where $G_0 = e^2/h$ is the conductance quantum, $e$ is the electron charge and $h$ is Planck's constant. The double degeneracy of the conductance channels stems from the spin-valley-locked bands at  $K$ and $K'$. The observed conductance quantization is expected and well understood as 1D ballistic transmission subbands. We note that in previous reports, 1D transport in monolayer and few-layer TMDs had plateau-like features separated by conductance values of $G_0$\cite{Pisoni2017, Marinov2017, Epping2018, Sakanashi2021, JBC2023}, a behaviour that was not fully understood. In addition to the expected 2$G_0$ degeneracy, we observe additional features at $G_0$ and $3G_0$ which are the focus of this paper (Fig.\,3a).

\begin{figure}[t]
    \centering
    \includegraphics[width=\textwidth]{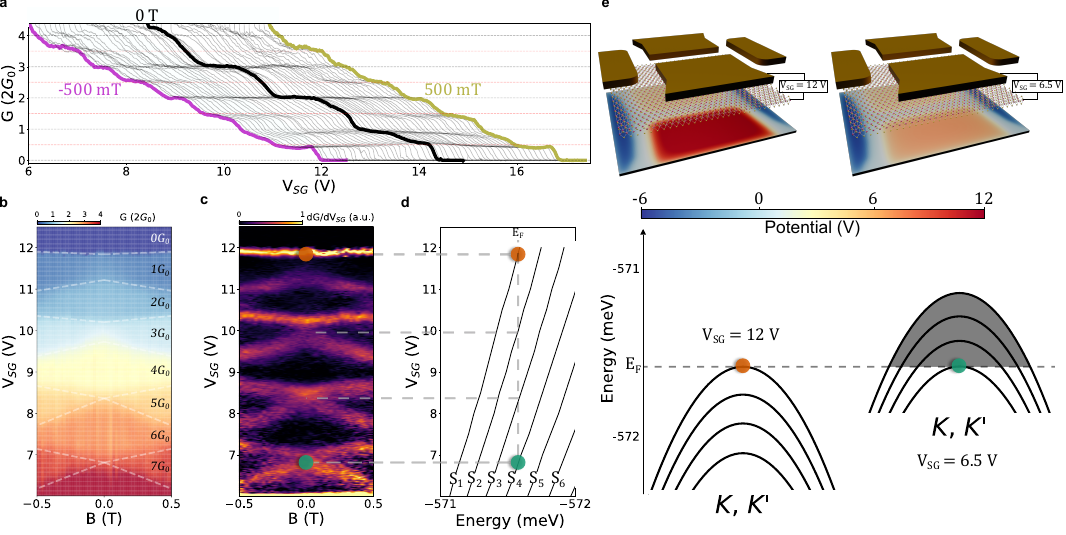}
    \caption{\textbf{Magneto-transport properties of a gate-defined 1D channel in monolayer WSe$_2$.} \textbf {(a)} Conductance across the 1D channel as a function of the voltage applied to the upper split gate and magnetic field. The magnetic field ranges from -500 mT (purple curve) to 500 mT (gold curve) with a 20 mT interval. The central black curve corresponds to B$_\perp$ = 0 T. The lower split gate, back-gate, and contact gates are all kept constant at 4 V, -3.5 V, and -7 V respectively. Individual curves are offset by 100 mV along the horizontal axis for clarity. \textbf {(b)} Color map of the conductance as a function of the upper split gate voltage and magnetic field. White dashed lines separate regions of different conductance. \textbf {(c)} Transconductance as a function of the upper split gate voltage and magnetic field obtained from the derivative of the data in (b). \textbf {(d)} Subband edge energies as a function of the upper split gate voltage computed with the anisotropic massive Dirac Fermion model. A black vertical dashed line shows the position of the Fermi energy. \textbf {(e)} The potential landscape (upper pannel) and resulting subband dispersions (lower pannel) when $V_{SG} = 12$ V and $V_{SG} = 6.5$ V. The position of the Fermi energy is indicated by the horizontal dashed line.}
\end{figure}

Applying $B_\perp$ between -500 mT  and 500 mT, corresponding to the purple and gold traces respectively in Fig.\,2a, the plateaus located at integer multiples of 2$G_0$ occurring at $B_\perp=0$ T evolve into plateaus with integer values of $G_0$ indicating that the two-fold degeneracy of the spin-valley locked bands has been lifted.  Figure 2b displays the conductance as a function of $B_\perp$ and $V_{SG}$ where regions of constant conductance are outlined by white dashed lines, demonstrating the robustness of the 1D subbands. To precisely chart the subband edges, Fig.\,2c presents the transconductance of the 1D channel by taking the numerical derivative dG/dV$_{SG}$. Dark regions indicate constant conductance and the bright lines represent the transition between two plateaus. Remarkably, a clear subband splitting is observed at magnetic fields as low as $\approx$ 250 mT. We also note that the additional $G_0$ and $3G_0$ plateaus observed at $B_{\perp} = 0$ T do not split with B$_\perp$, hinting towards the existence of a spin-valley polarized state at zero magnetic field.

To understand our observations, we develop an anisotropic massive Dirac Fermion model (see SI) described by the Hamiltonian
\begin{equation}
    \hat{H}_{MDF} = \frac{\Delta}{2} \sigma_z 
    + \alpha V(y) I_2
    + \frac{\lambda}{2} (I_2 - \sigma_z) 
    + \hbar v_F (\tau  k_x \sigma_x +
    \frac{v_{\perp}}{v_F} k_y \sigma_y), 
\end{equation}
where $\tau=\pm1$ for the $K$ ($K'$) valley,
$\Delta = 1.6$ eV is the main bandgap, 
$2\lambda = 0.46$ eV is the spin-orbit constant,
$\hbar v_F = 0.3927$ eV$\cdot$nm is the 
Fermi velocity, $I_2$ is the two-by-two identity matrix, and $\sigma_i$ ($i=x,y,z$) are the 
Pauli matrices.
The above effective 2D Hamiltonian is valid for both low-energy conduction and valence band states near the band edges at the $K$ and $K'$ valley and 
correctly accounts for the valley topology while including the bandgap and the spin-orbit interactions\cite{Peeters2018,Xiao2012,Szulakowska2019,Bieniek2018,Rose2013}.
The parameter $v_{\perp}$ denotes the effective 
Fermi velocity in the direction perpendicular to the channel.
Its value is a fitting parameter, and its departure from
the value $v_F$ is due to Coulomb interactions.
The channel potential $V(y)$ is multiplied by the 
lever arm parameter $\alpha = 1.2$ $\mu$eV/V (see SI).
We utilize this model to calculate the valence subband dispersions with $V(y)$ obtained by solving the Laplace equation numerically for all experimental values of V$_{SG}$ (see SI).

Figure 2d shows the evolution of the first few subband edge minima as a function of V$_{SG}$ at $B_\perp = 0$ T. Two examples of the potential landscape, when V$_{SG} = 12$ V and V$_{SG} = 6.5$ V, are plotted in the upper panel of Fig.\,2e with their respectively calculated subband dispersions in the lower panel. With the assumption that the Fermi energy in the leads does not depend on the gate potentials, we pin the Fermi energy at the onset of the first subband at V$_{SG} = 12$ V (red dot in Figs.\,2 c-e). By using the position of the intersection of the second subband and the Fermi energy as a fitting target for the effective Fermi velocity $v_{\perp}$, we find that the remaining calculated intersections agree well with the experiment (horizontal dashed lines in Figs. 2c,d).

This effective model captures well the conductance edges of doubly degenerate subbands, but it doesn't capture the features at $G_0$ and $3G_0$ at $B_{\perp} = 0$ T, highlighted by red arrows in Fig.\,3a. To do so, we include hole-hole interactions to the model using the Hartree-Fock (HF) approximation for two different scenarios of band filling as a function of the number of holes present in the transport channel (see SI). 

Our theoretical approach involves resolving the competition of single-particle and exchange energies close to the van Hove singularity characteristic for the density of states at the edge of one-dimensional subbands, which is smeared out as a result of the finite length of our channel. This approach was previously discussed in the context of the 0.7 anomaly, a partial spin polarization effect reported in semiconductors such as gallium arsenide, and whose origin is still under debate\cite{Pepper1996, Berggren1998,Cronenwett2002,Jaksch2006,Micolich2011,Ludwig2013,Micolich2013,Schimmel2017}.
First, we consider the lowest subband from each valley. The subbands may be populated with holes in a valley-symmetric configuration, schematically represented in Fig.\,3b, where each subband is filled with the same number of holes ($N/2$) and each contribute G$_0$ units of conductance. Alternatively, in the symmetry-broken configuration, only one subband originating from one valley is filled with $N$ holes leading to a total conductance of G$_0$ (Fig.\,3c). The corresponding energies $E_{HF}^{S}$ (HF energy of the symmetric configuration) and $E_{HF}^{SB}$ (HF energy of the symmetry-broken configuration) are calculated as:
\begin{equation}
    E_{HF}^{S}(N) = 2T(N/2) + U(N) - 2J(N/2).
    \label{hfene_symmetric}
\end{equation} 
\begin{equation}
    E_{HF}^{SB}(N) = T(N) + U(N) - J(N),
    \label{hfene_asymmetric}
\end{equation}
where $T$ is the total single-particle kinetic energy, $U$ is the total direct interaction energy, and $J$ the total exchange interaction energy.  We note that the direct term $U(N)$ is equal for both scenarios since it does not depend on the wave numbers of the subband states, only on their subband indices. Therefore the difference between these two total energies depends on the balance between the kinetic energy and the exchange energy. We find that these two energies depend differently on the total number of holes $N$. The total kinetic energy increases approximately quadratically with $N$, since the dispersion relation is approximately parabolic, whereas the total exchange energy increases quasi-linearly with $N$ due to the short-range nature of the exchange. As a result, we expect strong exchange interaction effects at small $N$, while at larger $N$, the difference between the energies $E_{HF}^{S}(N)$ and $E_{HF}^{SB}(N)$ will be mostly due to the difference in the total kinetic energy. The top panel of Fig.\,3d shows the HF energy  of the symmetric configuration (blue line) and the broken-symmetry configuration (orange line) ignoring the direct term. We find that for the first $\approx 100$ holes admitted into the system, the broken-symmetry configuration is lower in energy than the symmetric one. Thus, we find a spontaneous symmetry breaking brought about by the strong exchange interactions. However, as the total number of holes becomes larger than $100$, a SB-S transition occurs, and the symmetric configuration becomes the ground state. Consequently, only one conduction channel is available for the first $100$ holes, and the second conduction channel opens when the transition occurs, in qualitative agreement with experimental data.

\begin{figure}[t]
    \centering
    \includegraphics[width=\textwidth]{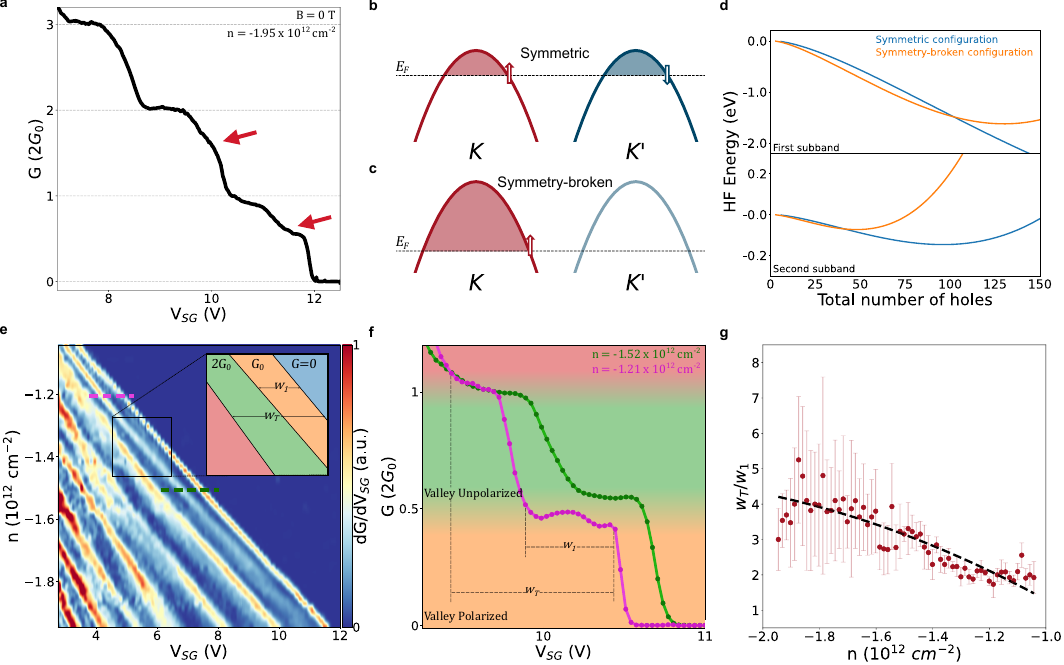}
    \caption{\textbf{Valley-polarized transport at zero magnetic field.} \textbf {(a)} Conduction as a function of V$_{SG}$ at B$_{\perp} = 0$ T where red arrows identify features at $e^2/h$ and 3 $e^2/h$. \textbf {(b)} \textbf {(c)} Depictions of two possible configurations for which holes can enter the system. In (b), the symmetric configuration, both valleys are populated and each contribute $G_0$ units of conductance, while in the symmetry-broken configuration (c), a single valley is filled with holes resulting in a total conductance of $G_0$. \textbf{(d)} Hartree-Fock energies of the symmetric configuration (blue) and symmetry-broken configuration (orange) as a function of the total number of holes for the lowest subband (top panel) and second subband (bottom panel). \textbf{(e)} Transconductance dG/dV$_{SG}$ as a function of the upper split-gate voltage V$_{SG}$ and carrier density in the leads as tuned by the back-gate voltage. The lower split gate and contact gates are kept constant at 4 V and -7 V respectively. \textbf{(f)} Conductance traces taken along the dashed lines in (e). \textbf{(g)} The ratio between the total width $w_T$ and the width of the SB spin-polarized plateau $w_1$ plotted as a function of carrier density. A quadratic trend is shown following the expected relation between the interaction strength $J$ and $n$.}
\end{figure}

A similar calculation can be performed for the second subband where the bottom panel of Fig.\,3d shows the relevant HF energies for the symmetric (blue line) and symmetry broken configuration (orange line). As for the first subband, the SB configuration is the ground state at a low subband population and transitions to the S configuration as more holes are introduced. This transition occurs at a lower number of holes in the subband stemming from the generally weaker exchange interactions in the second subband, and appears in experiment as a shoulder-like feature at 3G$_0$. Furthermore, due to the even weaker exchange interactions at higher subbands, no symmetry-broken configuration exists, in agreement with experimental results.

The effect of interactions can be tuned by changing the hole density $n$ of the WSe$_2$ with the back-gate. In Fig.\,3e, we plot the transconductance (dG/dV$_{SG}$) as a function of $n$ and V$_{SG}$ and observe that the widths of the conductance plateaus change with $n$. Focusing on the first two plateaus at $G_0$ and $2G_0$, Fig.\,3f shows traces taken at $n=-1.21\times10^{12}$ cm$^{-2}$ and at $n=-1.52\times10^{12}$ cm$^{-2}$ corresponding to the purple and green dashed lines in Fig.\,3e respectively, demonstrating the dependence of the width on $n$. We label the width of the SB spin-polarized plateau as $w_1$ and the total width corresponding to the sum of the SB spin-polarized plateau and the S spin-unpolarized as $w_T$. The ratio $w_T/w_1$ as a function of $n$ is plotted in Fig.\,3g where we confirm that $w_T/w_1 \propto J$, as predicted by our model. Calculated values for $J$ as a function of $n$ are fitted by a quadratic equation (see SI), and $w_T/w_1$ follows the same tendency scaled by constants (dashed black line in Fig.\,3g). This demonstrates that the hole-hole interactions are tunable in our device by the back-gate voltage.

\begin{figure}[t]
    \centering
    \includegraphics[width=\textwidth]{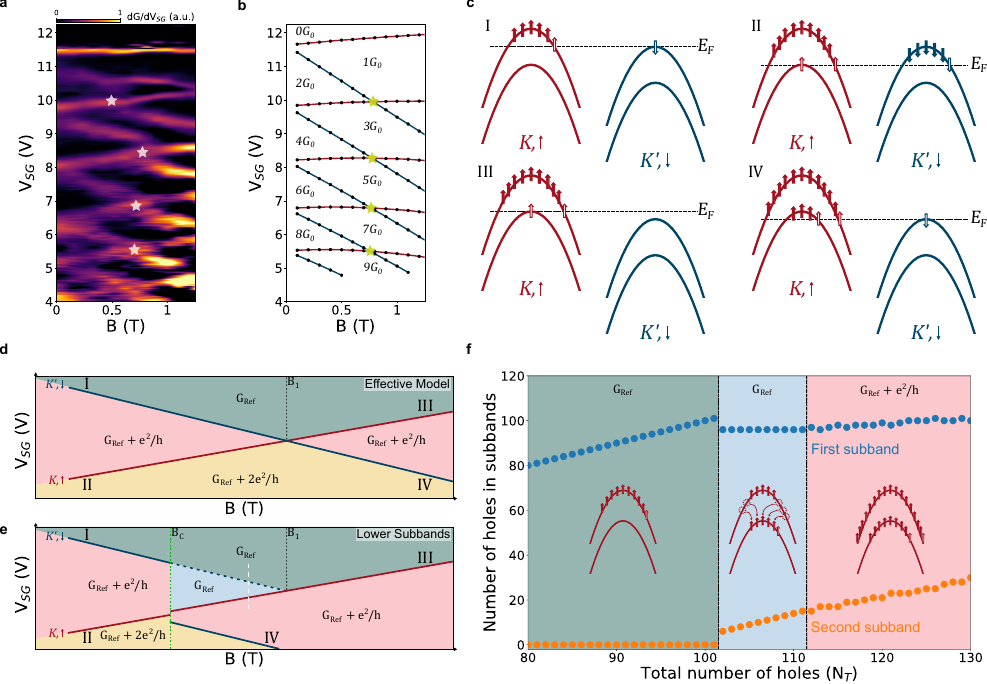}
    \caption{\textbf{Interaction driven reconstruction at high-magnetic field.} \textbf{(a)} 1D channel transconductance as a function of gate voltage and magnetic field. \textbf{(b)} Positions of the addition steps as a function of the split-gate voltage and the magnetic field calculated using the anisotropic massive Dirac Fermion mode and using $g_{e,Z} = -2$ and $g_{h,Z} = 2.6$. White and gold stars in (a) and (b) indicate where subband crossings occur. \textbf{(c)} Depictions of the possible single-particle subband configurations near a subband crossing. Filled arrows denote the holes already residing in the channel, and empty arrows denote the next holes being added to the system. \textbf{(d)(e)} Schematic of the conductance as a function of B$_{\perp}$ and V$_{SG}$ in the single particle picture (d) and when considering hole-hole interactions (e). The transition lines labelled by I - IV correspond to the configurations depicted in (c). The colored areas are labelled by their respective conductance values. \textbf{(f)} Occupation of the first two subbands at $K$ as a function of the total number of holes in the system obtained from Hartree-Fock calculations. This calculation is performed along the white dashed line in (e).}
\end{figure}

The application of a magnetic field beyond 0.5 T gives additional insights on the role of h-h interactions and how they depend on the subband population, as indicated in Fig.\,4a. White stars in Fig.\,4a indicate intervalley subband crossings where the spin-valley Zeeman splitting is equal to the intravalley subband spacing. We observe intersections with minute kinks for the third and higher crossings, however, clear discontinuities are evident for the first two\cite{Pepper2003, Pepper2004, Pepper2004_2}. To understand these observations, we include a magnetic field in the anisotropic massive Dirac Fermion model by adding to the Hamiltonian i) a Zeeman term $\frac{1}{2} g_{e(h),Z} \mu_B B$ where $g_{e,Z} = 2$ is the Landé factor of a free electron and $\mu_B$ is the Bohr magneton and ii) diamagnetic effects by replacing the momentum $\vec{k}$ with the generalized momentum $\vec{p_i}=\vec{k} + e\vec{A}$ where $e$ is the electron charge and $\vec{A}$ is the vector potential. The anisotropic hole Landé factor $g_{h,Z}$ is a fitting parameter. Choosing $g_{h,Z} = +2.6$ yields good agreement between theory and experiment as indicated in Fig.\,4b. Using this value of $g_{h,Z}$ we can fit the subband splitting in Fig.\,2c (or Fig.\,4a), and extract the hole valley g-factor $g_{h,v}$ for the first four subbands from $\Delta E = (g_{h,v}+g_{h,Z})\mu_B \Delta B$. We find that $g_{h,v}^1 = 13 \pm 3$, $g_{h,v}^2 = 21\pm 5$, $g_{h,v}^3 = 40  \pm 10$, and $g_{h,v}^4 = 40 \pm 10$. This suggests that as the 1D confinement and the interactions become stronger, the valley g-factor is reduced, distinct from 2D transport in monolayer TMDs\cite{Tutuc2017, Dean2018}. The reason behind this discrepancy is the 1D confinement potential. Indeed, hole-hole interactions enhance $g_{h,v}$ because of exchange while confinement weakens $g_{h,v}$ because it governs the energy quantization. In other words, the stronger the confinement is, the stronger the magnetic field is needed to form cyclotron orbits, leading to a smaller $g_{h,v}$. The extracted valley g-factors are a result of this competition.

We can qualitatively understand the nature of the crossings by using the HF approximation to calculate i) the total energies of the various subband filling configurations as a function of the magnetic field and ii) the number of holes in each subband as a function of the total number of holes for a fixed magnetic field. Figure 4c shows the four different configurations relevant in the system, while their corresponding transition lines in the B$_{\perp}$--V$_{SG}$ plane are schematically represented in Fig.\,4d. This representation corresponds to the single particle picture and does not include h-h interactions which is similar to the case of the third and higher subbands where interactions are less important. In the green region a certain number of subbands are already populated and contribute to the conduction, which we label as $G_{Ref}$. We  see that as $V_{SG}$ is decreased at low magnetic field, the subband filling follows configuration I and then configuration II, first filling the $K',\downarrow$ band with a step of $G_0$ and then the the second $K,\uparrow$ band with another $G_0$ conductance step. Above the crossover point $B_1$, the spin-valley Zeeman energy makes it energetically favorable to first fill  the $K,\uparrow$ band and then $K',\downarrow$ band. In experiment, for the lower subbands where interactions are stronger, we detect a discontinuity in the crossover region at a critical magnetic field labelled $B_c$. This effect is schematically shown in Fig.\,4e.  Following the white dashed line in this case, the system follows the subband filling sequence presented in Fig.\,4f, where we calculate the  energy of the first two subbands at $K$ as a function of the number of holes in the system. The first transition, between the green and blue regions, corresponds to a sudden redistribution of holes within the subbands in which  competition between kinetic and interaction energies transfers $\approx 5$ holes from the first to the second subband. This transition is not observed in our experiment as it only corresponds to a redistribution of holes and not to any additional conducting subbands, therefore there is no change in conductance. Within the blue region, the first subband population is frozen and further holes only fill the second subband. Transitioning to the pink region, it is energetically favorable to fill both bands leading to a $G_0$ increase in conductance. This sudden reallocation (transition from green to blue region) is due to the fact that the exchange energy is extremely local in $k$-space leading to a redistribution of the holes towards the vicinity of the subband minima, even though the promotion of the last added holes to the second subband costs kinetic energy. Furthermore, adding another hole to an already substantially filled lowest subband is not favourable energetically, since the new hole will be far
(in $k$-space) from the band minimum, whilst adding it to the slightly filled second subband brings it much closer to other holes. Thus, only one conduction channel exists, involving the second subband only. As that subband is filled with subsequent holes, they are added further away from the second subband minimum, and the advantage of the exchange energy becomes less pronounced. At this point, the holes are added to either subband in sequence, creating two conduction channels, which shows as a jump in conductance. This interaction-driven subband filling sequence leads to the particular discontinuities measured in Fig.\,4a for the lower subbands in which interactions are stronger.

We have demonstrated that a complete understanding of 
hole-hole interactions is necessary to explain all experimental features of the system. In systems of interacting fermions we typically deal with three aspects of interactions: direct, exchange, and correlations. In our theoretical model we utilized the Hartree-Fock approach, which accounts for the first two aspects, but underestimates the third. As we have demonstrated, the exchange interactions stabilize the spin-valley polarized phases. However, the inclusion of correlations lowers the energy of low-spin phases~\cite{Pawlowski2024}. Thus, we need to show that the spin-valley polarized, symmetry-broken phases remain ground states of our system even if the correlations are accounted for. We do this utilizing the configuration-interaction approach, which treats all three aspects of interactions on equal footing. However, its computational complexity makes it impossible to apply it to a system with the dimensions of this experiment, therefore we consider a one-dimensional channel with a width of $60$ nm, and length of $150$ nm. In Fig.\,5 we show the results of calculations of energies of six interacting holes at zero magnetic field. We find that the nature of the ground state of six holes strongly depends on the strength of Coulomb interactions, tuned by the choice of the dielectric constant $\varepsilon$. For a weakly interacting system [$\varepsilon=100$, Fig.\,5a], the ground state is the fully symmetric singlet configuration, with three holes in each valley. This state is separated from the excited states by a gap corresponding to the inter-subband spacing. With interactions of moderate strength
[$\varepsilon=20$, Fig.\,5b], the partially 
polarized states (color symbols) are separated from the 
unpolarized symmetric ground state by a much smaller gap.
Finally, with strong interactions [$\varepsilon=5$, close to experimental conditions, Fig.\,5c], the partially polarized, symmetry broken states and the symmetric singlet state are at similar energies, and a partially polarized state appears to take over as the global ground state of the system. We stress that in this strongly confined system the kinetic energy quantization is much stronger than in the channel probed experimentally, making it more difficult for the partially polarized states to stabilize.
Moreover, in this calculation the direct, exchange and correlations are screened by the same dielectric constant $\varepsilon$, while our experimental results demonstrate that the degree of screening depends on the aspect of interactions, resulting in the exchange being stronger than direct (see SI).
Thus, our symmetry broken, polarized ground state is found to be robust against correlations, even though the domains  of its stability in parameter space may be smaller than our Hartree-Fock approach predicts.

\begin{figure}[t]
    \centering
    \includegraphics[width=0.5\textwidth]{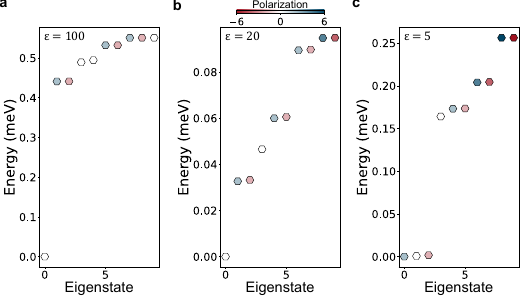}
    \caption{\textbf{Configuration-interaction picture.} Energies of six interacting holes at B$_{\perp} = 0$ T as a function of the lowest energy configurations. The color scale represents the spin-valley polarization of the configuration. The energies of the various configurations are calculated for different dielectric constants where $\epsilon = 100$ in panel a, $\epsilon = 20$ in panel b, $\epsilon = 5$ in panel c.}
\label{fig-CI}
\end{figure}

In conclusion, we elucidate the role that strong exchange interactions and  many body correlation effects play in quantum transport in TMDs and specifically how they affect the spin and valley degrees of freedom. The high quality of our device allows for ballistic transport in a monolayer WSe$_2$ 1D channel, resulting in several conductance quantization steps. We demonstrate that the degeneracy of each transport subband in this material is neither fourfold (containing both spin and valley degrees of freedom) nor single (as previously reported) but two-fold due to spin-valley locking. By tuning the carrier density, we control the interaction strength and can establish a fully spin-valley polarized transport channel at zero magnetic field with $G_0$ conductance. The origin of this spontaneous spin-valley polarization is a competition between kinetic and exchange energies leading to a breaking of the valley symmetry in the system. Using the HF approximation to include interactions, we demonstrate this symmetry breaking and furthermore predict discontinuities in the 1D magneto-transport spectrum which are corroborated by experiment. These findings demonstrate dramatic differences between TMDs and other quantum materials and can have deep consequences in the emerging field of valleytronics, where it is essential to understand the role of interactions between particles in different valleys to control the valley degree of freedom.

\section*{Methods}
\subsection*{Device fabrication}
Bulk crystals were exfoliated on a silicon substrate with 285 nm of thermal oxide to obtain the monolayer WSe$_2$, hBN, and few-layer graphite flakes used to fabricate the heterostructure illustrated in Fig.\,1 of the main text. Using an optical microscope, flakes with suitable thicknesses, geometry, and uniformity were identified. Their topographic properties were then confirmed using an atomic force microscopy (Bruker Dimension Icon) in tapping mode (SCANASYST-AIR). The heterostructure was assembled using a dry transfer technique aided by a polypropylene carbonate transfer polymer and operating with temperatures ranging from 40 $^\circ$C to 90 $^\circ$C. First, a bottom hBN flake measuring 24 nm in thickness was picked-up then used to pick-up a 4 nm thick flake of graphite. This stack of flakes was deposited on an intrinsic silicon substrate covered by 285 nm of thermally grown oxide. An electrical contact was made to the graphite flake by first employing a selective reactive ion etch using CF$_4$ and O$_2$ plasmas to etch a window in the hBN flake followed by metal deposition. Electron beam lithography and electron beam metal evaporation was used to pattern metallic contacts (2 nm of chromium + 8 nm of platinum) on top of the hBN. To remove any residue that may have accumulated on the surface of the contacts and the hBN during the fabrication process, the sample was annealed in a vacuum chamber (10$^{-7}$ Torr) at 300 $^\circ$C for 30 minutes. Additionally, an AFM operating in contact mode was used to displace any remaining residue residing on the contacts and hBN. A top hBN flake measuring 25 nm in thickness and monolayer WSe$_2$ were  sequentially picked-up, AFM cleaned, and dropped off onto the electrical contacts. After encapsulation, the heterostructure was once again annealed in a vacuum furnace following the same recipe as described earlier. The contact top gates ($V_{CG}$) and split gates ($V_{SG}$) were patterned using electron beam lithography and metallized using electron beam deposition. Additional optical micrographs of the device a various fabrication steps are provided in the Supplementary Information.

\subsection*{Experimental setup}
Transport measurements were performed in a Bluefors LD250 dry dilution refrigerator, equipped with an 8 T magnet, where measurements were conducted at the base temperature of 10 mK, unless otherwise noted. Standard four terminal DC conductance measurements were performed using a Keysight source measuring unit SMU2901a and an Agilent digital multimeter with a constant bias of 250 $\mu$V applied between the source and the drain contacts. All gates (split gates, contact gates, and back-gate) were independently tuned using individual Keysight source measuring units (SMU2901a/SMU2902b). See the supplementary information for the exact measurement configuration and individual contact characterization.

\section*{Data availability}
The data that support the findings of this study are available from the corresponding author upon reasonable request.



\section*{Acknowledgments}
We gratefully acknowledge Alexander Hamilton, Andrew Sachrajda, Guy Austing, and Jean Lapointe for their helpful discussions. This work was supported by the High Throughput and Secure Networks Challenge Program and the Quantum Sensors Challenge Program at the National Research Council of Canada. We acknowledge the support of the Natural Sciences and Engineering Research Council of Canada (NSERC) ALLRP/578466-2022, NSERC Discovery Grant No. RGPIN-2019-05714, University of Ottawa Research Chair in Quantum Theory of Quantum Materials, Nanostructures, and Devices and CIFAR. This research was enabled in part by support provided by the Digital Research Alliance of Canada (alliancecan.ca). This research was supported in part by PL-Grid Infrastructure. Data analysis and manuscript preparation (AL-M) was supported by the US Department of Energy, Office of Science, Basic Energy Sciences, Materials Sciences and Engineering Division under contract DE-AC02-06CH11357. JP acknowledges support from National Science Centre, Poland, under grant no. 2021/43/D/ST3/01989. K.W. and T.T. acknowledge support from the JSPS KAKENHI (Grant Numbers 21H05233 and 23H02052) and World Premier International Research Center Initiative (WPI), MEXT, Japan.

\section*{Competing interests}
The authors declare no competing financial or non-financial interests.

\section*{Author contributions}
J.B.-C., A.B., and L.G. designed the device architecture. P.B., and P.W. fabricated the top gate structure while J.B-C. fabricated the remainder of the device with inputs from A.L.-M. and L.G.. J.B.-C performed the experimental measurements and the data analysis with consultations from A.B., A.L.-M., and L.G..  M.K., J.P., D.M. and P.H. derived the theoretical models. K.W. and T.T. grew the hexagonal boron nitride crystals. J.B.-C, L.G. and M.K. wrote the manuscript with input from all authors. 

\graphicspath{}
\setcounter{figure}{0}
\setcounter{equation}{0}
\renewcommand{\figurename}{FIG.}
\renewcommand{\thefigure}{S\arabic{figure}}

\setlength{\parskip}{\medskipamount}

\title{Supplementary Information: Valley-spin polarization at zero magnetic field induced by strong hole-hole interactions in monolayer WSe$_2$}

\maketitle

\section{Device Fabrication and Contact Characterization}
\label{Device_Fabrication}

The contact resistance of individual contacts as a function of the contact gate voltage at 10 mK is obtained following the circuit diagram in Fig.\,S2a. The voltage probes of a 4-point measurement setup are placed on the contact of interest and an adjacent contact. Current is forced to flow from the contact of interest to a third contact with all other contacts floating. The measured resistance is therefore a sum of 1) the contact resistance of the contact connected to the voltage probe and the current source, and 2) a small contribution from the channel, therefore this technique slightly overestimates the contact resistance. The contact resistance is measured as a function of the contact gate directly above it, while the other contact gate is kept constant at -7 V to ensure all contacts in the circuit are activated. The back-gate is fixed at -3 V to activate the channel and reduce the contribution originating from channel resistance. The split-gates are both set to -0.5 V to avoid the formation of a 1D channel. Using this technique, the following contact resistances are extracted (Fig.\,S2 c-g): R$_{LC1}$ = 711 $\Omega$ and R$_{LC2}$ = 982 $\Omega$ at $V_{LCG}$ = -8 V, and R$_{RC1}$ = 5.31 k$\Omega$, R$_{RC2}$ = 4.55 k$\Omega$, and R$_{RC3}$ = 2.01 k$\Omega$ at $V_{RCG}$ = -7 V. For the 4-point measurements presented in the main manuscript, LC1 was connected to the current source, RC2 was connected to ground, and LC2 and RC3 were connected to the voltage meter. To further demonstrate the quality of the electrical contacts, Fig.\,S2h shows a linear relationship between the measured current and applied bias voltage between contacts LC1 and RC2 with a total resistance of 5.9 k$\Omega$ indicating that ohmic contacts are achieved.

\begin{figure}[ht]
    \centering
    \includegraphics[width=\textwidth]{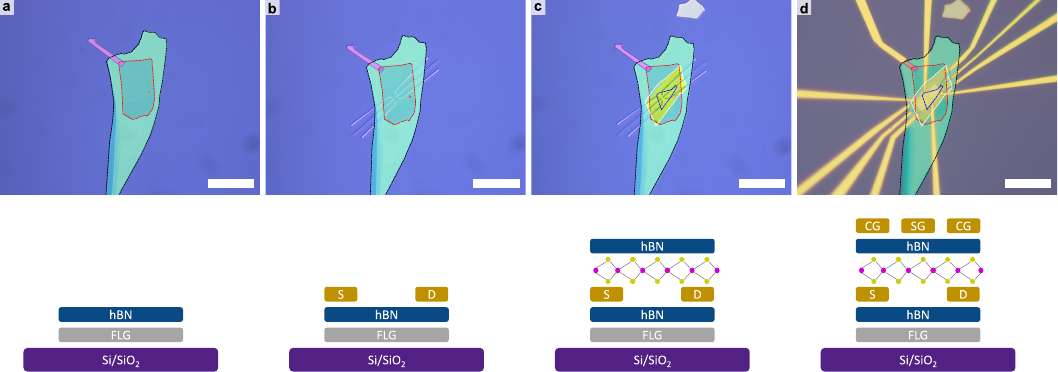}
    \caption{\textbf{Device fabrication.} \textbf{(a-d)} Optical micrographs (top) and cartoon schematics (bottom) showing the assembly process of the heterostructure device discussed in the main manuscript. In the optical micrographs, the individual flakes are outlined by the following colors: few-layer graphite = red, bottom hBN = black, monolayer WSe$_2$ = blue, and top hBN = red.}
\end{figure}

\begin{figure}[ht]
    \centering
    \includegraphics[width=\textwidth]{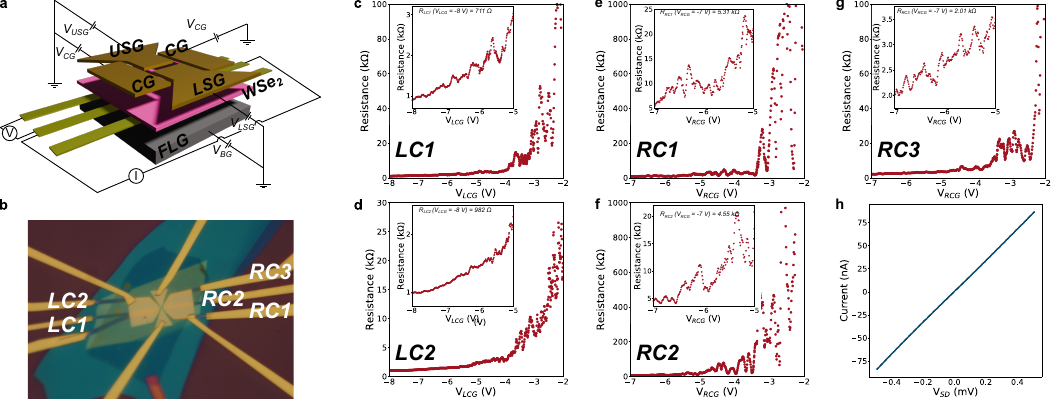}
    \caption{\textbf{Contact resistance and quality.} \textbf{(a)} Schematic illustration of the device studied in the main manuscript with an example of the circuit diagram used to measure the contact resistance of contact LC1. \textbf{(b)} Optical micrograph of the device where individual contacts are labelled. \textbf{(c-g)} Contact resistances of each contact as a function of the contact gates measured at 10 mK. When measuring the contact resistance of a contact located on the left (right) side, the right (left) contact gate is fixed at -7 V. V$_{BG}$ = -3 V and V$_{LSG}$ = V$_{USG}$ = -0.5 V. The inset focus on the low resistance range. \textbf{(h)} 2-point IV measurement using LC2 as the source and RC2 as the drain.}
\end{figure}


\section{Theoretical Model}
\subsection{Massive Dirac Fermion Model at Zero Magnetic Field}
\subsubsection{Hamiltonian and parametrization}

To describe our system and understand our observations, we begin with the massive Dirac Fermion model, in which the Hamiltonian is written in the two-band basis as
\begin{equation}
\hat{H}_{\tau} = \left[
    \begin{array}{cc}
    \frac{\Delta}{2} + \alpha V(y) & \hbar v_F (\tau \hat{k}_x - i \hat{k}_y ) \\
    \hbar v_F (\tau \hat{k}_x + i \hat{k}_y ) & -\frac{\Delta}{2} + \lambda + \alpha V(y)
    \end{array}
    \right].
    \label{hamil_MDF}
\end{equation}
Here, $\tau$ is the valley index, so that $\tau=1$ corresponds to the valley K, and $\tau=-1$ denotes the valley K'. The parameter $\Delta$ defines the bandgap, while $\lambda$ corrects the top of the valence band due to the spin-orbit interaction. The parameter $v_F$ is the Fermi velocity and $\hbar$ is the Dirac's constant. Further, $V(y)$ is the channel potential (discussed in SI III), and $\alpha$ is the lever arm translating the voltages into the energy scale (discussed in SI IV). Lastly, $\hat{k}_x$ and $\hat{k}_y$ are the components of the gradient operator, $\hat{\vec{k}} = -i \vec{\nabla}$.

The above Hamiltonian is formulated in the electron edge, i.e., it accounts for the coupling of the electronic states in the conduction band and the electronic states in the valence band. It is consistent with that given in Ref.~\onlinecite{Peeters2018}. Here we account for the spin-valley locking phenomenon and consider only two bands per valley, corresponding to the spin subspace forming the valence band edge. Specifically, we take only the spin-up subspace in the valley K, and the spin-down subspace in the valley K'. This is why in our treatment the spin-orbit parameter $\lambda$ modifies the top of the valence band identically in each valley. The parameters for WSe$_2$ are also taken from Ref.~\onlinecite{Peeters2018} and are $\Delta=1.6$ eV, $\lambda=0.23$ eV, and $\hbar v_F=0.3927$ eV$\cdot$nm.

\subsubsection {Computational procedure}
We assume that the one-dimensional channel is sufficiently long for us to treat it as translationally invariant along the $x$ axis, while the channel confinement $V(y)$ generated by the gates has to be taken into account explicitely along the $y$ axis. Since $V(y)$ is in a numerical form, we look for the eigenenergies and eigenstates of the Hamiltonian (\ref{hamil_MDF}) in a numerical computational procedure. To this end, we choose the following basis set:
\begin{equation}
    \langle \vec{r}|n,k\rangle = 
    \psi_{n,k}(x,y) = \frac{1}{\sqrt{L}} \exp\left( i k x \right)
    \sqrt{\frac{2}{W}} \sin\left[ \frac{n\pi}{W} \left(y+\frac{W}{2}\right) \right].
\end{equation}
These basis functions are products of the function in the $x$ direction (with translational symmetry) and the $y$ direction (the confinement direction).

In the $x$ direction we choose a plane wave function reflecting the translational invariance. Further, we assume periodic boundary conditions on our one-dimensional channel, which require
\begin{equation}
    \exp(i k 0) = \exp(i k L),
\end{equation}
and thus discretize the wave vector $k= 2\pi m / L$, with the integer $m=0,\pm1, \pm2,\dots$. We note that the Hamiltonian (\ref{hamil_MDF}) conserves the wave vector $k$.

In the $y$ direction we choose to enclose our potential $V(y)$ in an infinite quantum well of width $W$. Our basis of sines is simply the basis of eigenstates of that quantum well, with the origin chosen in the middle of the channel. The basis functions are enumerated by the integer $n=1,2,\dots$. We take the width $W=563.2$ nm, slightly narrower than the spatial extent of the calculated potential profiles (Sup. III). This choice of $W$ is made by finding the minimal region encompassing the spatial variation of the potential, i.e, beyond which the potential becomes flat (coordinate-independent).

As the MDF model is a two-band approach, we seek the single-particle eigenstates in the dimer form
\begin{equation}
    |k, S\rangle = \left[ 
    \begin{array}{c}
    \sum_{n=1}^{N_B} A_n^{k,S} |n,k\rangle \\
    \sum_{n=1}^{N_B} B_n^{k,S} |n,k\rangle  
    \end{array}
    \right],
    \label{sp_state}
\end{equation}
where $S$ is the subband index, the coefficients $A_n^{k,S}$, $B_n^{k,S}$ form the eigenvectors of the Hamiltonian (\ref{hamil_MDF}), and $N_B$ is the basis size (we take N$_B=150$). The above eigenstates, as well as the corresponding single-particle energies $E_{S,k}$ are obtained by formulating the Hamiltonian (\ref{hamil_MDF}) in a matrix form in our basis and diagonalizing it numerically.

\subsubsection {Single-hole states in the channel}
\label{sp_no_interactions}
Figure S3a shows the dispersion calculated for the gate potential generated with the gate voltage $V_{USG}=12$ V. We focus on the valence band states only. The single-particle energies are plotted as a function of the wave number $k$, i.e., the index of the plane wave along the channel. In the hole language the energy increases towards the bottom of the graph, and therefore the topmost trace corresponds to the lowest hole subband confined in the channel, $S=1$. As we go downwards in energy, we encounter the subsequent subbands, $S=2, 3, \dots$. We find that the dispersions for the valleys K and K' are exactly degenerate.

\begin{figure}[ht]
    \centering
    \includegraphics[width=0.75\textwidth]{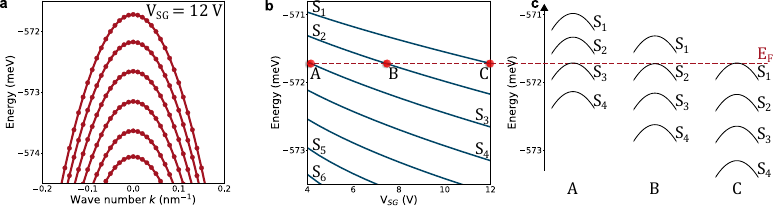}
    \caption{\textbf{Single-particle Dispersion.} \textbf{(a)} Dispersion for the one-dimensional channel with the confinement potential calculated for $V_{USG} = 12$ V. \textbf{(b)} Evolution of the edges of several lowest hole subbands as a function of the gate voltage $V_{USG}$. \textbf{c} Sketch of the band dispersion alignment at points A,B, and C. The red dashed line marks the Fermi energy.}
\end{figure}

Assuming that the Fermi energy in the leads do not depend on the gate potentials, as the potential $V_{USG}$ is made less positive, the hole subbands shift and cross the fixed Fermi level one by one. From Fig.\,S3a it is evident that this crossing involves the bottom of the subband, i.e., the state at $k=0$, because this is the lowest-energy hole state within each subband. In the choice of energy axis consistent with Fig.\,S3a, gradual decrease of the voltage $V_{USG}$ will result in a shift of all the subbands upwards in energy. In that evolution, whenever the Fermi energy aligns with the bottom of any new subband, a new conduction channel opens, resulting in a step in conductance.

Figure S3b shows the evolution of several of the lowest channel subband edges (i.e., subband energies at $k=0$) as a function of the voltage V$_{USG}$. The red dashed line shows the Fermi energy, which we have chosen to reproduce the first conductance step occurring at V$_{USG}$ = 12 V. This is consistent with the experimental data shown for $B=0$. We note that these subband edges are doubly degenerate owing to the valley degeneracy. Therefore, at V$_{USG}$ = 12 V (diagram C in Fig.\,S3c) we would expect an onset of conductance of {\em two} units, i.e., by $2e^2/h$, because two conduction channels become simultaneously available for the holes. The next onset, by another {\em two} units, would occur at the voltage V$_{USG}\approx$ 7.5 V (diagram B in Fig.\,S3c). The third step, again by {\em two} units, is expected at V$_{USG}\approx$ 4 V (diagram A in Fig.\,S3c). This is, however, not what we see experimentally. It is evident from experiment that the first conductance step, occurring at V$_{USG}$ = 12 V, is only by {\em one} unit of $e^2/h$. The second step follows at V$_{USG}\approx$ 11.2 V and is also by one unit of conductance. The third and fourth steps, one unit each, occur at V$_{USG}\approx$ 10.3 V and $9.8$ V, respectively. The fifth and sixth signature occurs at V$_{USG}\approx$ 8.3 V and $6.9$ V, respectively, and at $B=0$ each of these steps is by $2e^2/h$.

We therefore draw two conclusions: first, a mechanism exists for removing the degeneracy of the two lowest pairs of subbands, which has been explained at length in the main manuscript, and second, our single-particle model appears to overestimate the gaps between subsequent subbands, causing them to cross the Fermi energy for voltages V$_{USG}$ much less positive than those recorded in experiment. Therefore, we now turn to including the interactions in our model approach.

\subsection{Hartree-Fock Model of Interacting Holes in the One-Dimensional Channel}
\subsubsection{The Hamiltonian}

We now account for the Coulomb interactions among the holes populating the single-particle levels. Let us denote the creation (annihilation) operator of a hole on the level $\gamma=\{S,k,\sigma\}$ by $c^+_{\gamma}$ ($c_{\gamma}$).  In our composite orbital index $\gamma$ we have added the spin-valley index $\sigma$ to the quantum numbers $S, k$ describing the single-particle orbital. We note that we only need one spin-valley index owing to the spin-valley locking, as explained in the discussion of the Hamiltonian (\ref{hamil_MDF}). In consequence, $\sigma=\uparrow$ will denote a spin-up hole in the valley K, and $\sigma=\downarrow$ will denote a spin-down hole in the valley K'.

In the language of the creation and annihilation operators, the Hamiltonian of interacting holes in our one-dimensional channel is written as

\begin{equation}
    \hat{H}_{I} = \sum_{\gamma} E_{\gamma} c^+_{\gamma} c_{\gamma}
    + \frac{1}{2} \sum_{\gamma_1,\gamma_2,\gamma_3,\gamma_4}
    \langle \gamma_1, \gamma_2 |V| \gamma_3, \gamma_4\rangle
    c^+_{\gamma_1}c^+_{\gamma_2}c_{\gamma_3}c_{\gamma_4},
\end{equation}
where $\langle \gamma_1, \gamma_2 |V| \gamma_3, \gamma_4\rangle$ denote the matrix elements of the Coulomb interaction calculated in the basis of single-particle orbitals.

\subsubsection {The Hartree-Fock approach}
To understand a system containing many holes, we employ a simplified Hartree-Fock (HF) model, where we calculate the total energies (expectation values of the Hamiltonian $\hat{H}_I$, or HF energies) of several candidate hole configurations. The ground state of the system of $N$ holes is identified as the configuration with the lowest HF energy. 

The HF treatment formally starts with writing down a candidate configuration in the form of a single Slater determinant. In the most general terms, we assume that we populate different subbands in both valleys with holes up to a certain state (indexed by momentum $k^{(S,\sigma)}_{MAX}$), which depends on the subband index and the valley. Our configurations therefore takes the following form

\begin{equation}
    |\Psi\rangle = \prod_{S=1}^{S^{\uparrow}_{MAX}} 
    \prod_{k=-k^{(S,\uparrow)}_{MAX}}
    ^{k^{(S,\uparrow)}_{MAX}}    
    c^+_{S,k,\uparrow}
    \prod_{S'=1}^{S^{\downarrow}_{MAX}} 
    \prod_{k=-k^{(S,\downarrow)}_{MAX}}^{k^{(S,\downarrow)}_{MAX}}
    c^+_{S',k,\downarrow}    
    |0\rangle,
\end{equation}

where $|0\rangle$ is the vacuum state. If we traverse the set of all {\em occupied} configurations ${S,k,\sigma}$ by the index $\gamma$, the HF energy corresponding to the above configuration is

\begin{equation}
    E_{HF}(\Psi) = \langle \Psi | \hat{H}_I | \Psi\rangle 
    = \sum_{\gamma} E_{\gamma} 
    + \frac{1}{2} \left[ \sum_{\gamma,\gamma'} 
    \langle \gamma, \gamma' | V | \gamma', \gamma\rangle
    - \langle \gamma, \gamma' | V | \gamma, \gamma'\rangle \right].
    \label{total_HF}
\end{equation}

In this energy, the first term is the total single-particle energy of the holes. The second term accounts for direct and exchange interactions, represented respectively by the first and second Coulomb element within the sum.

The addition energy of the $(N+1)$$^\textrm{st}$ hole to a system of $N$ interacting holes, which occupy the states from the set $\{ \gamma \}$, will depend on the index $\gamma_1 = \{S_1,k_1,\sigma_1\}$ of the single-particle orbital to which that hole is added. This addition energy is equivalent to the HF quasiparticle energy for the orbital $\gamma_1$ in the presence of the $N$ holes and can be expressed as

\begin{equation}
    E_{HF}(\gamma_1) = E_{\gamma_1}
    + \sum_{\gamma} 
    \langle \gamma, \gamma_1 | V | \gamma_1, \gamma\rangle
    - \langle \gamma, \gamma_1 | V | \gamma, \gamma_1\rangle.
\end{equation}

The addition energy of the hole added to the orbital $\gamma_1$ will consist of the single-particle energy of that orbital (the first term) and the selfenergy $\Sigma(\gamma_1)$, which accounts for all repulsive direct interaction elements with the $N$ resident holes (the second term), and all attractive exchange interaction terms with these holes (the third term).

\subsubsection {Coulomb matrix elements}
\label{model_coulomb}
To complete the HF model, we have to specify the Coulomb direct and exchange matrix elements. Since we have the single-particle orbitals in the form of Eq. (\ref{sp_state}), we can obtain these matrix elements by direct integration of the Coulomb potential. However, the single-particle basis set $N_B=150$ makes it prohibitively expensive to apply a full numerical calculation of a set of matrix elements required for tens of holes. This is the case even if we account for the fact that in Eq. (\ref{sp_state}) most of the coefficients $A$ are negligibly small, leaving us only with large amplitudes $B$ for half of that state. Instead, we notice that the hole subbands have a nearly parabolic dispersion close to the band minima, as is evident from Fig.\,S3a, and the channel confinement potential (Sup.~\ref{Channel_Potential}) is also approximately parabolic near to its bottom. We will therefore approximate the hole wave functions by

\begin{equation}
    \langle \vec{r} | k, S \rangle =  
    \frac{1}{\sqrt{L}} \exp(ikx) f_{S}(y),
\end{equation}

where the subband functions $f_{S}$ are taken in the form of the eigenstates of a one-dimensional harmonic oscillator. These functions are scaled by a distance parameter $l$, i.e., the oscillator length. This length is usually expressed by the particle effective mass and the frequency of the harmonic confinement. In our approach we will extract the values of $l$ from the channel confinement profiles by noticing that at the distance $l$ from the origin, the value of the harmonic potential energy is equal to the ground-state energy above the potential floor (it is a classical turning point). We note further that the wave number $k$ in our single-particle functions is measured from the valley momentum, $K$ and $-K$ for the valley K and K', respectively. This will be important in distinguishing between the intravalley and intervalley matrix elements.

The direct matrix element calculated with the $1/r$ Coulomb potential takes the form

\begin{equation}
    \langle \gamma, \gamma_1 | V | \gamma_1, \gamma \rangle =
    \frac{e^2}{4\pi\epsilon_0 \epsilon_r L^2 } \int_0^L dx_1 \int_0^L dx_2
    \int_{-\infty}^{\infty} dy_1 \int_{-\infty}^{\infty} dy_2
    \frac{|f_{S}(y_1)|^2|f_{S_1}(y_2)|^2}{\sqrt{(x_1-x_2)^2+(y_1-y_2)^2}},
    \label{direct_interaction}
\end{equation}

with $e$, $\epsilon_0$, and $\epsilon_r$ being the electron charge, the vacuum electrical permittivity, and the dielectric constant of the material, respectively. The subband index $S$ ($S_1$) originates from the compound index $\gamma$ ($\gamma_1$). We find that the direct term depends exclusively on the subband indices. It does not depend on the wave numbers $k$, $k_1$ of the particles involved. Moreover, for the same subband indices, the values of the intervalley and intravalley elements are identical.

As for the intravalley exchange element, we have

\begin{eqnarray}
    \langle \gamma, \gamma_1 | V | \gamma, \gamma_1 \rangle &=&
    \frac{e^2}{4\pi\epsilon_0 \epsilon_r L^2 } \int_0^L dx_1 \int_0^L dx_2 \nonumber\\
    &\times& \int_{-\infty}^{\infty} dy_1 f_{S}(y_1)f_{S_1}(y_1) 
    \int_{-\infty}^{\infty} dy_2 f_{S}(y_2)f_{S_1}(y_2)
    \frac{\exp\left[i(k-k_1)(x_1-x_2)\right]}{\sqrt{(x_1-x_2)^2+(y_1-y_2)^2}}.
\end{eqnarray}

This element depends on both the subband indices and the wave numbers of the states involved. The intervalley exchange element is zero because of the spin-valley locking: holes in opposite valleys have opposite spins, which breaks the selection rules for the exchange interaction.

All matrix elements are scaled by the dielectric constant $\varepsilon_r$ appropriate for our sample. Optical studies of multi-layer WSe$_2$ samples reveal $\varepsilon_r\approx 6.24$ (Ref.~\onlinecite{Hou2022}), while density-functional calculations predict $\varepsilon_r\approx 7.2$ for a monolayer and $\varepsilon_r\approx 8.1$ for a bulk WSe$_2$ material~\cite{Laturia2018}. Furthermore, both experimental~\cite{Kang2020} and theoretical~\cite{Laturia2018,Bieniek2022} treatments show that $\varepsilon_r$ depends not only on the number of monolayers, but also on the material encapsulating the sample (in our case, hBN). In Ref.~\onlinecite{Bieniek2022} it is shown that for hBN-encapsulated monolayer systems, it is actually more correct to assume $\varepsilon_r$ equal to that of the dielectric, in our case the value of $3.5$. As is evident, it is not clear which value is the correct one, particularly in conditions where nearby gates and holes in the two-dimensional gas in the leads may screen some aspects of the Coulomb interactions. As a result, we will treat $\varepsilon_r$ as a fitting parameter and discuss in detail the dependence of our results on its specific value.

\subsection{Addition Energies of Holes in the Interacting System and the Anisotropic Massive Dirac Fermion Model}

\begin{figure}[ht]
    \centering
    \includegraphics[width=0.5\textwidth]{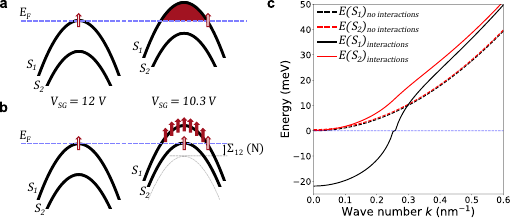}
    \caption{\textbf{Effect of Coulomb interactions on the subband dispersion.} Alignment of the hole subbands in the K valley (K' valley behaves identically) at gate voltages $V_{USG}=12$ V (left) and $V_{USG} = 10.3$ V (right) based on \textbf{(a)} the single-particle model without interactions and \textbf{(b)} model accounting for Coulomb interactions. \textbf{(c)} Energies of the two lowest subbands as a function of the wave number $k$. Dashed lines show the bands calculated with the single-particle Hamiltonian without interactions for the gate voltage $V_{USG}=10.5$ V. Solid lines show the addition energies, i.e., the subband energies renormalized by direct and exchange interactions with $2N=126$ resident holes, $N$ holes per valley. Blue dashed lines show the Fermi energy.}
\end{figure}

\subsubsection{Schematic Picture}

In Fig.\,S4a we show schematically the alignment of hole subbands for the noninteracting model for the gate voltage $V_{USG}=12$ V (left) and $V_{USG}=10.3$ V (right). These subbands are plotted for the valley K, with their K' equivalents behaving identically. These voltages are chosen as they correspond to the onset of the lowest subband and the second subband, respectively (with the proper account for the valleys). In our noninteracting model, we have chosen the Fermi energy (represented by the blue line in Fig.\,S4a to coincide with the lowest hole subband at $V_{USG}=12$ V. This accounts for the first conductance step, corresponding to the addition of the first hole, denoted schematically by the hollow red arrow, into the channel. However, when considering the Massive Dirac Fermion model, at the gate voltage $V_{USG}=10.3$ V our second subband is not low enough in energy to coincide with the Fermi energy. The step heralding the opening of the second subband for conduction is expected only at $V_{USG}=7.5$ V, as is evident from Fig.\,S3b.

Figure S4b shows the analogous subband alignment for the case accounting for interactions. The left panel is identical to the case in (a), because we are adding the first hole to the system (hollow red arrow), and there are no other holes to interact with. On the other hand, for the voltage $V_{USG}=10.3$ V, we will now expect the conductance step, i.e., a hole being successfully added onto the second subband (hollow red arrow). Thus, the {\em quasiparticle} second subband (the solid line) will now be lowered in energy relative to the {\em noninteracting} second subband (dashed line) by the selfenergy $\Sigma_{12}(N)$, describing the interaction of the new hole with all $N$ holes (red arrows) already present in the lowest subband. As is evident from our HF model, the holes being in the same valley and having the same spin will interact via exchange, which is attractive (lowers the quasiparticle energy). This mechanism allows us to find a quantitative correspondence of our simple model with the experimental data.

\subsubsection{Calculation of the selfenergy}

We assume that the lowest subband has been filled with $2N$ holes, $N$ holes per valley, in the symmetric configuration. Indeed, in the experimental data we see that the onset of the second subband appears for voltages significantly lower than the phase transition from the polarized to the unpolarized lowest subband. As explained before, this transition occurs at $V_{USG}=11.1$ V, i.e., the conduction step from $e^2/h$ to $2e^2/h$. The selfenergy appearing in the addition energy of the hole to the second subband in valley K, denoted in Fig.\,S4a by $\Sigma_{12}(0)$, will therefore take the following form:

\begin{equation}
    \Sigma_{12}(N) = \sum_{\gamma} \left[ 
        2 \langle \gamma, \gamma_1 | V | \gamma_1, \gamma\rangle
    - \langle \gamma, \gamma_1 | V | \gamma, \gamma_1\rangle
    \right],
\end{equation}

where $\gamma = \{ S=1,k=2\pi m/L, \sigma=\uparrow  \}$ and $\gamma_1 = \{ S_1=2,k_1=0, \sigma_1=\uparrow \}$, and the first-subband wave vector $m$ changes from $-m_F$ to $m_F$, that is, runs over all holes occupying the first subband symmetrically. The maximum index is related to the number of holes $N$ in each valley by $N=2m_F+1$. The index $m_1$ of the second subband is zero, as we are adding the new hole to the bottom of the second subband. We note that there is a factor of two in front of the direct term. This accounts for the equal filling of both valleys and the fact that, as shown by Eq.~(\ref{direct_interaction}), the value of the direct Coulomb matrix element does not depend on the wave numbers of the two electrons. Let us denote a single direct term by $U_{12}$. This term is obtained from Eq.~(\ref{direct_interaction}) by setting $S=1$ and $S_1=2$. As a result, the entire sum over direct elements evaluates to $2NU_{12}$. The sum over the exchange elements, on the other hand, is carried out only in one valley because of the spin selection rule. Utilizing the harmonic oscillator functions for our two subbands, we arrive at the selfenergy in the following form:

\begin{eqnarray}
    \Sigma_{12}(N) &=& 2NU_{12} \nonumber - V^J_{12}(N),\\
    V^J_{12}(N)&=& \frac{e^2}{4\pi\epsilon_0\epsilon_r L} \left[  \frac{4\pi^2l^2}{L^2}\right]
        \sum_{m_1=-m_F}^{m_F} 
        \exp\left(  \frac{2\pi^2 m_1^2l^2}{L^2} \right)
        \int_{m_1}^{\infty} dv \sqrt{v^2-m_1^2}
        \exp\left( - \frac{2\pi^2l^2}{L^2} v^2 \right). \nonumber
\end{eqnarray}

The total direct contribution is proportional to the total number of holes. On the other hand, on closer inspection, we find that the sum $V^J_{12}(N)$ of exchange elements converges to a fixed value as $m_F$ is increased (i.e., as the number $N$ of holes in the lowest subband increases). This reflects the local nature of the exchange interaction.

\subsubsection{Interaction effects in intersubband gaps}

Evidently, in order to be able to calculate $\Sigma_{12}(N)$, we need to know the number of holes occupying our system at the gate voltage $V_{USG}=10.3$ V. This number is not known a priori, but we can estimate it in the following way. Our point of departure is the fact that the addition of a hole into the channel happens only if the addition energy of that hole is equal to the Fermi energy of the leads. The first hole is added to the lowest subband at the gate voltage $V_{USG}=V_1=12$ V. In this case, since there are no other holes in the channel, the addition energy of that hole is simply equal to its single-particle energy. Therefore, the Fermi energy of the leads $E_F = E_{S=1}(k=0,V_1)$. Now, we assume that $E_F$, being defined by the many-hole state in the leads, does not depend on the gate voltage $V_{USG}$. As the gate voltage is adjusted to $V_{USG}=V_2=10.3$ V, the hole is now added to the second subband, and so its addition energy must be equal to $E_F$ at that point. Therefore, we can write 

\begin{equation}
    E_{S=2}(k=0,V_2) + 2NU_{12} - V^J_{12}(N) = E_{S=1}(k=0,V_1).
    \label{calibration_1}
\end{equation}

We emphasize that the single-particle energy $E_{S=2}(k=0)$ is extracted at the voltage $V_{USG}=V_2=10.3$ V, while $E_{S=1}(k=0)$ is extracted at $V_{USG}=V_1=12$ V. In the absence of interactions, the above equation trivially simplifies to the equality of single-particle energies at the two voltages. This is easily understood: as the gate voltage $V_{USG}$ is made less positive, the single-particle energies descend and align with $E_F$ one by one, opening new conduction channels. However, from our single-particle model we have $E_{S=2}(k=0,V_2) - E_{S=1}(k=0,V_1) = 0.3$ meV, that is, not zero. Therefore, the interaction component appearing on the left-hand side of Eq. (\ref{calibration_1}) must be negative to compensate for this mismatch in single-particle energies. In consequence, the total exchange $V^J_{12}(N)$ must be larger than the total direct term $2NU_{12}$. The numerical calculations of Coulomb elements require the length parameter, which for $V_{USG}=10.3$ V is $l=24.5$ nm. Assuming the dielectric constant $\varepsilon_r=3.5$, we find $U_{12}=3.45$ meV and $V^J_{12}(N)$ dependent nontrivially on $N$. For example, for $N=11$, $V^J_{12}(N) = 4.837$ meV, for $N=101$, $V^J_{12}(N) = 17.486$ meV, for $N=201$, $V^J_{12}(N) = 22.057$ meV, and for $N=301$, $V^J_{12}(N) = 24.756$ meV. Evidently, the total direct term $2NU_{12}$ is orders of magnitude larger than the total exchange term, and the negative interaction correction required in Eq. (\ref{calibration_1}) simply cannot be realized for any number $N$ due to the excessively large value of $U_{12}$. If this were true in the experimental system, the addition to the second subband would never take place, since making $V_{USG}$ even less positive (i.e., adding more holes into the lowest channel subband) only makes the alignment worse.

To address this discrepancy, we include additional screening of the direct interactions, while the exchange interactions remain screened only by the dielectric constant $\varepsilon_r$ of the material. This screening originates from the nearby gates which maintain the holes inside the channel. Of course, we have to expect that the holes present in the system must not repel too strongly, otherwise we would not be able to see the essentially ballistic signatures of transport, characteristic for systems with near to zero repulsion. Therefore, we postulate that the direct elements have to be additionally scaled by a constant $\varepsilon_S$ accounting for the screening. This choice, however, introduces another unknown into Eq.~(\ref{calibration_1}) which requires a second equation to be able to find both $N$ and $\varepsilon_S$.

The second equation can be formulated by looking again at the bottom-right hand panel of Fig.\,S4b. Up to now we have tracked the hole added to the bottom of the second subband, denoted by the hollow red arrow in the subband $S=2$. However, the conduction channel through the lowest subband remains open: a hole can be added to the level just above the Fermi energy in the subband $S=1$, as denoted by a hollow red arrow at the extreme right of the panel. Evidently, the addition energy of either hole must be the same:

\begin{equation}
    E_{S=2}(k=0,V_2) + 2NU_{12} - V^J_{12}(N) = E_{S=1}(m=m_F+1,V_2) + 2NU_{11} - V^J_{11}(N).
    \label{calibration_2}
\end{equation}

The left-hand side of this equation is the addition energy of the hole to the bottom of the second subband, while the right-hand side represents the addition energy of the hole to the edge of the filled first subband (one orbital above the Fermi momentum $k_F=2\pi m_F / L$ in that subband). Here, the single-particle energy $E_{S=1}(m=m_F+1,V_2)$ corresponds to the hole added to the first available state in the subband $S=1$ and is calculated at the gate voltage $V_{USG}=V_2=10.3$ V, just as the second-subband energy $E_{S=2}(k=0,V_2)$. The direct element in the lowest subband is $U_{11}=3.95$ meV if the extra screening is not included. The exchange term at the edge of the lowest subband is

\begin{equation}
    V^J_{11}(N) = \sum_{\gamma} 
    \langle \gamma, \gamma_1 | V | \gamma, \gamma_1\rangle,
\end{equation}

where $\gamma = \{ S=1,k=2\pi m/L, \sigma=\uparrow  \}$ and $\gamma_1 = \{ S_1=1,k_1=2\pi (m_F+1) / L, \sigma_1=\uparrow \}$. The summation extends over the lowest-subband wave vectors from the negative Fermi momentum $-k_F$  to the positive Fermi momentum $k_F$. The new hole, referred to by the composite index $\gamma_1$, is placed on the first available single-particle state, i.e., one with the momentum $k_1= 2\pi (m_F+1) / L$. The Fermi momentum is defined by the number $N$ of holes per valley by $2m_F +1 = N$. We find that, similarly to the second subband, the exchange at the edge of the lowest subband depends on the number $N$ of holes in the valley, but converges as $N$ increases. For $\varepsilon_r=3.5$ we find the following values: for $N=11$, $V^J_{11}(N) = 12.331$ meV, for $N=101$, $V^J_{11}(N) = 26.634$ meV, for $N=201$, $V^J_{11}(N) = 31.225$ meV, and for $N=301$, $V^J_{11}(N) = 33.924$ meV. Compared to the exchange $V^J_{12}(N)$ at the bottom of the second subband, the values of $V^J_{11}(N)$ are significantly larger, particularly for lower values of $N$.

We introduce explicitly the effective screening of the direct term and rewrite our two equations in the following form:

\begin{eqnarray}
    E_{S=1}(k=0,V_1) - \left[E_{S=2}(k=0,V_2) - V^J_{12}(N)  \right]  &=& 
     2N\frac{1}{\varepsilon_S}U_{12}, \label{exec1} \\
    \left[E_{S=2}(k=0,V_2) - V^J_{12}(N)\right] 
    - \left[E_{S=1}(k=k_F+1,V_2) - V^J_{11}(N) \right]  
    &=&   2N\frac{1}{\varepsilon_S}U_{12}\left(\frac{U_{11}}{U_{12}} - 1 \right).
    \label{exec2}
\end{eqnarray}

We divide these two equations sidewise and end up with one equation, which does not depend on $\varepsilon_S$ and depends on the number of holes $N$ only implicitly (through the exchange and single-particle energies). The direct terms enter only as the ratio $U_{11}/U_{12} = 1.145$. We find that our equation is solved for $2N=130$ holes, i.e., $65$ holes per valley. This number does not contradict our earlier description of symmetry-broken states, as it puts us firmly in the regime of the symmetric lowest-subband configuration, which becomes the ground state configuration of the lowest subband at $2N=102$ holes. We insert the number of holes into Eq.~(\ref{exec1}) and solve for the effective screening parameter, obtaining $\varepsilon_S = 31.35$. We find, within our model, that the direct term is strongly screened. This explains essentially ballistic transport spectra seen experimentally in spite of the fact that the channel contains hundreds of holes.

We map out the dependence of the effective screening constant $\varepsilon_S$ on the overall strength of interactions, quantified by the dielectric constant $\varepsilon_r$. In the above analysis, we took $\varepsilon_r=3.5$, and the effective screening $\varepsilon_S$ acted on the direct interaction term on top of $\varepsilon_r$, while the exchange interaction was scaled by $\varepsilon_r$ only. Using the same approach, we have computed $\varepsilon_S$ and the critical number $2N$ of holes corresponding to the onset of the second subband for several model values of the dielectric constant $\varepsilon_r$. For $\varepsilon_r=4.0$, we find $2N=122$ and $\varepsilon_S=30.39$. Further, for $\varepsilon_r=5.5$, we find $2N=106$ and $\varepsilon_S=28.51$; for $\varepsilon_r=7.0$, we find $2N=96$ and $\varepsilon_S=27.15$, and for $\varepsilon_r=8.5$, we find $2N=88$ and $\varepsilon_S=25.94$. In general, as the interactions are made weaker, the critical number $2N$ of holes becomes smaller. In all cases, however, the onset of the transport involving the second subband occurs for a larger total number of holes than that corresponding to the phase transition to the symmetric configuration in the lowest subband. We find, therefore, no contradiction to the experimental data over a broad range of the system parameters. Furthermore, the effective screening parameter $\varepsilon_S$ of the direct interactions appears to depend relatively weakly on the dielectric constant $\varepsilon_r$. Indeed, changing $\varepsilon_r$ from $3.5$ to $8.5$, i.e., more than doubling it, leads to the decrease of $\varepsilon_S$ only by about $17$\%. This is because the effective screening depends more on the ratios of different Coulomb elements rather than on their values, although, as is evident in Eq.~(\ref{exec1}), this dependence involves also the single-particle energies and is therefore nontrivial. A more significant change of this parameter would be expected if we tuned the system so that the onsets of the first and second subbands correspond to altogether different gate voltages. Indeed, that would lead to different intersubband spacing, as well as different channel confining potentials, which would lead to a different length parameter $l$ determining the Coulomb matrix elements. Therefore, the relative robustness of $\varepsilon_S$ against tuning $\varepsilon_r$, but not gate voltages, suggests that this extra screening indeed originates from the gates, and would most likely be dependent on the geometry of the sample as well as the overall density of holes in the leads, all of these characteristics being gate-tunable.

\subsubsection{Effective Single-Particle Model, the Anisotropic Massive Dirac Hamiltonian}

As can be seen from the previous section, the interactions strongly modify the addition energies of the system. In particular, from Fig.\,S3a we see that the single-particle intersubband gaps are of order of $0.3$ meV, the single-particle dispersion can span several meV, while the exchange corrections can be several times that value. To illustrate this, in Fig.\,S4c we show the energies of the two lowest channel subbands, $S_1$ (black) and $S_2$ (red) without interactions (dashed lines) and with interactions accounted for (solid lines). For the case without interactions, we simply plot $E_{S=1}(k)$ and $E_{S=2}(k)$ choosing the origin so that $E_{S=1}(0)=0$. For the case with interactions, the system is filled with $2N=130$ holes, in the symmetric configuration with $N=65$ holes per valley. The blue dashed line denotes the Fermi energy corresponding to this occupation. We plot the quasiparticle energies:

\begin{eqnarray}
    E^{HF}_{S=1}(k) &=& E_{S=1}(k) + \frac{N+N^*}{\varepsilon_S}U_{11} \nonumber\\
        &-& \sum_{\substack{k_1=-k_F \\ k_1\neq k}}^{kF} 
        \langle S=1,k_1,\uparrow; S=1, k, \uparrow | V | 
        S=1,k_1,\uparrow; S=1, k, \uparrow \rangle, \\
    E^{HF}_{S=2}(k) &=& E_{S=2}(k)+ \frac{2N}{\varepsilon_S}U_{12} \nonumber\\
    &-& \sum_{k_1=-k_F}^{kF} 
    \langle S=1,k_1,\uparrow; S=2, k, \uparrow | V | 
        S=1,k_1,\uparrow; S=2, k, \uparrow \rangle.
\end{eqnarray}

The number $N^*$ in the expression for $E^{HF}_{S=1}(k)$ equals $N-1$ when we are calculating the energy for $k\leq k_F$ (in the interior of the hole droplet), and $N$ otherwise. This removes the selfinteraction effects in the direct interaction.

We see that the dispersion of the two lowest subbands is substantially changed by the interaction effects. Without interactions, the two lowest subbands are separated by a gap of order of $0.44$ meV, which is approximately constant for all wave numbers $k$. Their quasi-parabolic dispersion can be characterized by the effective mass $m^*=0.37$ $m_0$, with $m_0$ being the free electron mass. On the other hand, the interactions open a large intersubband energy gap, of order of $22$ meV, at $k=0$. This gap strongly depends on the wave number: it decreases only slightly as a function of $k$ in the interior of the droplet but decreases rapidly at its edge. The lower subband appears to be much more affected by the hole selfenergy than the upper one. Upon further increase of $k$, the two subbands approach each other. This characteristic behavior is caused mainly by the large difference in exchange energies for the lowest and second subbands. Moreover, the exchange being local, its renormalization of the quasiparticle energies decreases rapidly as we explore the quasiparticle energies further away from the Fermi energy. For large values of $k$, the two quasiparticle subbands are shifted from the ones for the noninteracting system by the direct term.

It is clear that this behavior of quasiparticle energies cannot be reproduced in detail by the simple massive Dirac Fermion model described by the Hamiltonian~(\ref{hamil_MDF}). However, in reproducing the experimental results, we are only interested in the energy of the band edges, i.e., the bottom of each subband at $k=0$. Moreover, the signatures of the addition steps appear when the holes begin to occupy the next unoccupied subband, as we discussed in the previous Section. Remarkably, we have found that a very good reproduction of the experimental results can be achieved by utilizing the anisotropic massive Dirac Fermion model presented in the main manuscript, with the Hamiltonian (\ref{hamil_MDF}) modified as follows:

\begin{equation}
    \hat{H}^{(eff)}_{\tau} = \left[
        \begin{array}{cc}
        \frac{\Delta}{2} + \alpha V(y) & \hbar v_F \tau \hat{k}_x - i \hbar v_F^{(eff)} \hat{k}_y  \\
        \hbar v_F \tau \hat{k}_x + i \hbar v_F^{(eff)} \hat{k}_y  & -\frac{\Delta}{2} + \lambda + \alpha V(y)
        \end{array}
        \right].
        \label{hamil_MDF_eff}
\end{equation}

Here, $v_F^{(eff)}$ is the effective Fermi velocity in the $y$ direction, that is, in the direction perpendicular to the channel. By tuning this parameter we account for the overall shift of the subband edge energy due to the direct and exchange contributions. We note that the dispersion along the channel is unchanged, which corresponds well to the approximately unchanged dispersion of the quasiparticle subbands close to the subband edge visible in Fig.\,S4b. We have achieved an excellent fit with the experimental data by choosing $\hbar v_F^{(eff)}=0.15$ eV$\cdot$nm, i.e., a value lower than the original one by a factor of $2.62$. We assume, as previously, that the onset of the conductivity involving the first subband takes place at the gate voltage $V_{USG}=12$ V which gives us the Fermi energy of the leads. The Fermi energy intersects the subsequent subbands at voltages corresponding very well to all the complete subband steps observed experimentally, as seen in Figs.\,2(b-d) of the main text. Of course, without the inclusion of interactions, the single-particle model does not reproduce the symmetry breaking effects, and the subband onsets corresponds to the conductance steps of $2e^2/h$. While we match the model to the first conductance peak at $V_{USG}=12$ V, the onset of the second subband is reproduced by our model to be between the single-height steps close to $V_{USG}=10$ V. We stress that this effective model is set up only to reproduce the experimental peak positions, but does not reveal the microscopic nature of the shifts of subband edges relative to the bare single-particle model.

\subsection{Symmetry Breaking for the Lowest Two Subbands}

In this section, we expand the discussion of the symmetry breaking event, primarily focusing on the explanation of the terms found in equations 2 and 3 of the main text.

The first term in equations 2 and 3 of the main text is the total single-particle (kinetic) energy. Suppose that we have $N$ holes in valley K. The states in that valley will be occupied up to a certain index $m_M$ such that $2m_M+1=N$, and their total kinetic energy can be expressed as 

\begin{equation}
    T(N) = E_{S=1}(m=0) + 2 \sum_{m=1}^{m_M} E_{S=1}(m)
\end{equation}

owing to the fact that the subband dispersion is symmetric. The energies $E_{S=1}(m)$ are calculated as appropriate eigenenergies of the Hamiltonian~(\ref{hamil_MDF}).

Next, we compute the total direct interaction energy. The direct terms do not depend on the wave numbers of the subband states, only on their subband indices. Since we distribute all the holes on the first subband, the direct interaction of each pair is expressed by the same matrix element $U_{11}$. Therefore, irrespective of their distribution, the total direct repulsion energy of $N$ holes will be 

\begin{equation}
    U(N) =  \frac{N(N-1)}{2} U_{11}.
\end{equation}

Finally, we compute the total exchange interaction. For $N$ holes in the valley K, this component of the total energy is expressed as

\begin{equation}
    J(N) = \frac{1}{2}\sum_{m_1=-m_M}^{m_M} \sum_{\substack{m_2=-m_M \\ m_2\neq m_1}}^{m_M} 
    \langle S=1,m_1,\uparrow; S=1,m_2,\uparrow | V |S=1,m_1,\uparrow; S=1,m_2,\uparrow \rangle.
\end{equation}

The exchange elements for the subbands $S=1$ are obtained in a closed form. However, to obtain their numerical values, we need to establish the oscillator length $l$ appropriate for our channel confinement. For the gate voltage $V_{USG}=12$ V, we extract $l\approx 24$ nm. Furthermore, we assume that in the range of voltages $V_{USG}=11$ to $12$ V, the potential profile changes very little, and we can utilize the parameters and data generated for $V_{USG}=12$ V throughout.

Now we formulate the total energies of the two configurations (equations 2 and 3 of the main text), $E_{HF}^{S}(N)$ for the symmetric one, and $E_{HF}^{A}(N)$ for the one with broken time-reversal symmetry. In the former case, we have $N/2$ holes in each valley, and 

\begin{equation}
    E_{HF}^{S}(N) = 2T(N/2) + U(N) - 2J(N/2).
    \label{hfene_symmetric}
\end{equation}

In the latter, all $N$ holes are in one valley, and

\begin{equation}
    E_{HF}^{SB}(N) = T(N) + U(N) - J(N).
    \label{hfene_asymmetric}
\end{equation}

We find that each configuration has an identical direct energy $U(N)$, and we can disregard it. Therefore, the difference between these two total energies depends on the balance of the kinetic and exchange energies. We find that these two energies depend differently on the total number of holes $N$. The total kinetic energy increases approximately quadratically with $N$, since the dispersion is approximately parabolic, with the effective mass $m^* = 0.37$ $m_0$ (with $m_0$ being the free electron mass). For a small $N$, $T(N)$ is in the $\mu$eV range, while for $N$ large enough it attains the values in tens of meV. On the other hand, the exchange interaction of two holes on neighboring lowest-subband states is of order of $2.5$ meV. However, the total exchange energy $J(N)$ increases with $N$ quasi-linearly due to the short-range nature of the exchange. As a result, we expect strong exchange interaction effects at small $N$, while at larger $N$ the difference between the energies $E_{HF}^{S}(N)$ and $E_{HF}^{SB}(N)$ will be mostly due to the difference in the total kinetic energy. This is what is seen and explained in the main text.

We have analyzed the stability of the spin and valley-polarized phase against the symmetrical one for the values of $\varepsilon_r=4.0$, $5.5$, $7.0$, and $8.5$, the latter being already larger than the theoretical estimates for the WSe$_2$ bulk dielectric constant. In all cases we find the symmetry-broken, valley-polarized phase to be the ground state for a small number of holes. The different values of $\varepsilon_r$ only influence the critical number of holes for which the transition to the valley-symmetric configuration takes place. We find this number to be $N_C = 96$ for $\varepsilon_r=4.0$, $N_C = 80$ for $\varepsilon_r=5.5$,   $N_C = 70$ for $\varepsilon_r=7.0$, and $N_C = 64$ for $\varepsilon_r=8.5$, for the first subband. Our HF analysis shows, therefore, that the symmetry-broken valley polarized phase is robustly stable over a broad range of system parameters.

Let us now turn to the splitting of the second subband. We will discuss two configurations analogous to those described by Eqs. (\ref{hfene_symmetric}) and (\ref{hfene_asymmetric}), however the holes will be distributed both on the first ($S=1$) and second ($S=2$) subbands. When the signatures of the second subband appear in transport (at the gate voltage $V_{USG}\approx 10$ V), there are sufficiently many holes in the system for the occupation of the first subband to have transitioned to the symmetric configuration. Moreover, we assume that the number of holes in the first subband is much larger than that in the second subband, which just begins to be populated at these gate voltages. The exchange energy experienced by the hole on the second-subband state $\gamma_2 = \{S=2,k_2,\uparrow\}$ from all the holes on the lowest subband states $\gamma_1 = \{S=1,k_1,\uparrow\}$ is expressed as 
\begin{equation}
    V_J^{2-1}(k_2) = \sum_{\gamma_2} \langle \gamma_1, \gamma_2 |V|\gamma_1, \gamma_2\rangle.
\end{equation}
Now we account for the fact that the exchange interaction depends on the difference $k_1-k_2$ of the wave numbers of single-particle states, and decreases rapidly with the increase of $|k_1-k_2|$. As a result, for a very large first-subband occupation, the above energy does not depend on the index $k_2$. In other words, the holes on the first subband form a translationally invariant continuum, which corrects the energy of each second-subband state by the same amount. This approximation allows us to exclude the exchange term $V_J^{2-1}(k_2)$ (i.e., the interaction with the first subband) from the total HF energy of second-subband configurations. Thus, in comparing the energies of the symmetric and polarized second-subband configurations, we can use formulas analogous to Eqs.~(\ref{hfene_symmetric}) and~\ref{hfene_asymmetric}), respectively, only with the kinetic and exchange energy specialized to the second subband. We perform our calculations by building the total kinetic energy with the dispersion calculated for the voltage $V_{USG}=10$ V, and the total exchange energy with the length parameter $l = 24.5$ nm, appropriate for the region of interest.

As in the case of the lowest subband, we have analyzed the stability of the valley-polarized phase against the strength of interactions, quantified by the value of the dielectric constant $\varepsilon_r$. Again, adjusting this parameter influences the critical number $N_C$ of electrons corresponding to the phase transition to the valley-symmetric configuration. We find this number to be $N_C = 38$ for $\varepsilon_r=4.0$, $N_C = 32$ for $\varepsilon_r=5.5$,   $N_C = 28$ for $\varepsilon_r=7.0$, and $N_C = 24$ for $\varepsilon_r=8.5$. We again find that, in our HF analysis, the valley-polarized phase is robustly stable in our parameter space, however its stability region systematically shrinks as the interactions are made weaker.

\section{Numerical Solutions of the Channel Potential}
\label{Channel_Potential}
We simulate the potential landscape at the level of the monolayer WSe$_2$ by solving numerically the Laplace equation with boundary conditions corresponding to the chosen gate voltages, and assuming that the electric field approaches zero at a very large distance below the gate (von Neumann boundary conditions). Using the methodology described in details in Refs. 1 and 2, and using the calculated dielectric constant of the hBN (3.5) under the gates, we generate the electrostatic potential $V(x,y)$ at the level of the WSe$_2$ flake. The cross-sections of $V(x,y)$ extracted along the red dashed line in Fig.\,S5a are plotted in Fig S5b for a fixed voltage of 4 V on the lower split gate and  a varying voltage from 4 V to 12 V applied to the upper split gate. The complete potential landscape $V(x,y)$ is plotted when the upper split gate is set to 6.5 V (Fig.\,S5c) and 12 V (Fig.\,S5d) while the lower split gate is kept at 4 V.

\begin{figure}[ht]
    \centering
    \includegraphics[width=\textwidth]{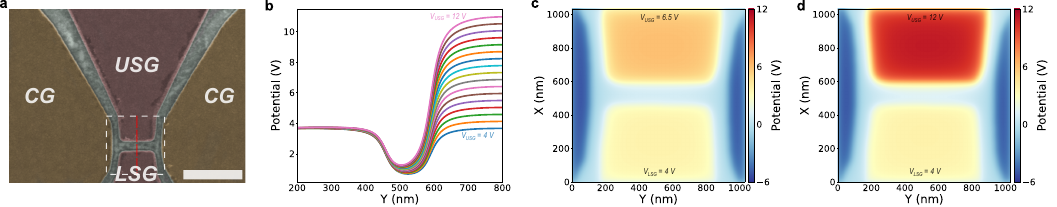}
    \caption{\textbf{Potential Landscape.} \textbf{a)} False-color scanning electron micrograph of the top gate structure of the device. The scale bar corresponds to 1 $\mu$m. \textbf{b)} Profiles of the electrostatic potential at the level of the WSe$_2$ flake calculated along the red dashed line (a). The lower split gate is fixed at 4 V while the upper split gate varies from 4 V to 12 V at a 0.5 V interval. \textbf{c-d} Colormap of the electrostatic potential at the level of the WSe$_2$ flake of the white dashed area in (a). The lower split gate is fixed at 4 V while the upper split gate is fixed at 6.5 V (c) and 12 V (d).}
\end{figure}

\section{Finite Bias Spectroscopy and $\alpha$-parameter extraction}
\label{Bias_Spectroscopy}
To obtain the lever arm of our device, a source-drain voltage, applied between contacts LC1 and RC2, is swept and the current traversing the device is measured. To eliminate the contribution from contact resistances, we monitor the voltage drop between contacts LC2 and RC3. The result is plotted in Fig.\,S6a as a function of the upper split gate where V$_{Bias}$ is the voltage measured between LC2 and RC3. The conductance is obtained (Fig.\,S6b) by taking a numerical derivative with respect to the bias voltage. A second derivative, this time with respect to the split gate voltage, is performed to obtain the transconductance (Fig.\,S6c). From this plot, characteristic diamonds are visible (dashed black lines) where their slope corresponds to the lever arm:

\begin{equation*}
\alpha = \dfrac{dV_{Bias}}{dV_{SG}} = 0.0012 \pm 0.0003
\end{equation*}

\begin{figure}[ht]
    \centering
    \includegraphics[width=0.72\textwidth]{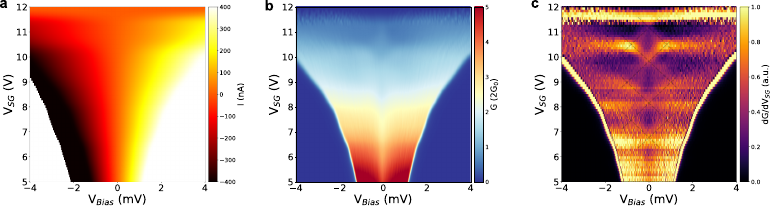}
    \caption{\textbf{Finite Bias Spectroscopy of the 1D Channel.} \textbf{(a)} Current traversing the device as a function of the split gate voltage and the measured voltage across the channel. \textbf{(b)} Numerical derivatives with respect to the bias voltage and \textbf{(c)} with respect to the split gate voltage. Black dashed lines outline the characteristic diamonds.}
\end{figure}

\section{Exchange Interaction scaling vs. Carrier Density}

The Coulomb exchange matrix element is calculated as a function of the split gate-voltage and is shown to be well fitted by a quadratic function (Fig.\,S7a). From Fig.\,3e of the main text, a clear linear relationship is found between the carrier density and the split-gate voltage. From these two tendencies, we find that the Coulomb exchange matrix element is related to the carrier density in the following form (dashed line in Fig.\,3g of the main text):

\begin{equation*}
J  = -0.0011\left(\dfrac{n[10^{12}\, \textrm{cm}^{-2}]+0.66}{-0.11}\right)^2+0.038\left(\dfrac{n[10^{12}\, \textrm{cm}^{-2}]+0.66}{-0.11}\right)+2.2
\end{equation*}

To observe this relation experimentally, we measure and plot the ratio $w_T/w_1$ as a function of the carrier density (Fig.\,3g of the main text). The total width of the plateau $w_T$ measured as a split gate voltage can be related to the energy spacing between the first and second subband of a single valley:

\begin{equation*}
w_T  = \alpha\Delta E_{2-1}
\end{equation*}

The energy spacing between the two subbands can be approximated as the energy spacing between two harmonic oscillator energy levels, which is related to the oscillator length $l$ as:

\begin{equation*}
w_T \propto \dfrac{\alpha \hbar}{ml^2}
\end{equation*}

The width $w_1$ directly probes the number of holes in the system at the transition between the SB-configuration and the S-configuration. We calculate this number of holes theoretically and find that it is linearly related to the inverse of the oscillator length (Fig.\,S7b), therefore:

\begin{equation*}
w_1 \propto \dfrac{1}{l}
\end{equation*}

Furthermore, calculations show that the oscillator length is inversely proportional to the Coulomb exchange matrix element (Fig.\,S7c), therefore we expect the ratio $w_T/w_1$ to be proportional to $J$.

\begin{figure}[ht]
    \centering
    \includegraphics[width=0.85\textwidth]{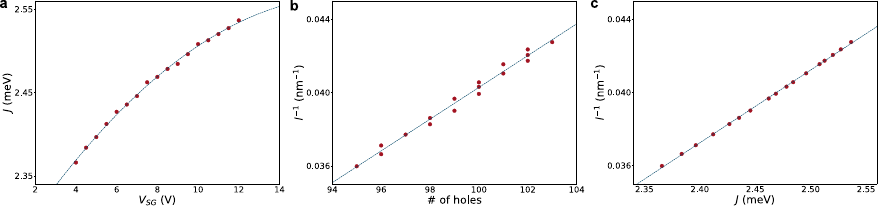}
    \caption{\textbf{Scaling of the Coulomb exchange matrix element} Red data points represent calculated values for (a) the Coulomb exchange matrix element as a function of the upper split gate voltage, (b) the inverse of the oscillator length as a function of the number of holes at the transition, and (c) the inverse of the oscillator length as a function of the Coulomb exchange matrix element.}
\end{figure}



\end{document}